\documentclass[aps,pra,showpacs,citeautoscript,amsmath,amssymb,floatfix,superscriptaddress,bbm, verbatim,amsfonts,changes,10pt,nofootinbib,longbibliography]{revtex4-1}
\pdfoutput=1
\usepackage{amsmath,bbold,bm}
\usepackage{amssymb}
\usepackage{epsfig}
\usepackage{xcolor}
\usepackage{amsthm}
\usepackage{appendix}
\usepackage{cancel}
\usepackage{stmaryrd}
\usepackage{soul}
\usepackage{ulem}
\usepackage{todonotes}
\usepackage{multirow}
\usepackage{hyperref}
\usepackage{xifthen}

\def\<{\langle}
\def\>{\rangle}

\newcommand{\be}{\begin{eqnarray} \begin{aligned}}
\newcommand{\ee}{\end{aligned} \end{eqnarray} }
\newcommand{\benn}{\begin{eqnarray*} \begin{aligned}}
		\newcommand{\eenn}{\end{aligned} \end{eqnarray*} }
\newcommand{\ben}{\begin{eqnarray} \begin{aligned}}
\newcommand{\een}{\end{aligned} \end{eqnarray} }

\newcommand{\bc}{\begin{center}}
	\newcommand{\ec}{\end{center}}

\newcommand{\id}{\mathbb{1}}

\newcommand{\tr}{\mathop{\mathsf{tr}}\nolimits}

\newcommand{\beq}{\begin{eqnarray} \begin{aligned}}
\newcommand{\eeq}{\end{aligned} \end{eqnarray} }
\newcommand{\bea}{\begin{array}}
	\newcommand{\eea}{\end{array}}
\newcommand{\bee}{\begin{enumerate}}
	\newcommand{\eee}{\end{enumerate}}
\newcommand{\bei}{\begin{itemize}}
	\newcommand{\eei}{\end{itemize}}
\newtheorem{theorem}{Theorem}

\newcommand{\ag}{{\boldsymbol\alpha}}
\newcommand{\bg}{{\boldsymbol\beta}}

\usepackage{amsfonts}

\def\01{\{0,1\}}

\newcommand{\ket}[1]{|#1\rangle}
\newcommand{\bra}[1]{\langle#1|}

\newcommand{\proj}[1]{|#1\rangle\!\langle#1|}

\renewcommand{\sout}[1]{\ignorespaces}
\newcommand{\foothide}[1]{\ignorespaces} 
\newcommand{\jono}[1]{\textcolor{black}{#1}}

\def\<{\langle}
\def\>{\rangle}

\def\s{\,\,\,\,}

\newtheorem*{rep@theorem}{\rep@title}
\newcommand{\newreptheorem}[2]{
	\newenvironment{rep#1}[1]{
		\def\rep@title{#2 \ref{##1} (restatement)}
		\begin{rep@theorem}}
		{\end{rep@theorem}}}
\makeatother
\newreptheorem{thm}{Theorem}
\newreptheorem{lem}{Lemma}

\def\p{{\bf p}_{\vec k}}

\def\dtau{{\tau}} \DeclareMathAlphabet\mathbfcal{OMS}{cmsy}{b}{n}

\def\X{\Lambda}
\def\T{\hat{T}}
\def\z{{z}}

\def\NE{\hat{N}}
\def\L{{\hat{L}}}

\def\Hq{\hat{H}}
\def\Hqm{\hat{H}^{(m)}}
\def\hc{{h}}
\def\H{{\cal H}}

\def\baseab{e^{\alpha\beta}}
\def\PQdiv{\boldsymbol{\nabla}_{\alpha\beta}}

\def\dist{{\Delta}}
\def\ddist{{d\!\dist}}
\def\ddf{{\mathcal D}\dist}

\def\u{u}
\def\g{{g}}
\def\gfdet{\sqrt{g}}

\def\Tnnc{\ham}

\def\ham{{\cal H}^{(m)}}
\def\superhamtot{{\cal H}^{(tot)}}
\def\qsuperhamtot{\hat{\mathcal{H}}^{(tot)}}
\def\superhamgrav{{\cal H}^{(gr)}}
\def\supermomgrav{{\cal H}^{(gr)}}
\def\supermomtot{{\cal H}^{(tot)}}
\def\qsupermomtot{\qsuperhamtot}%
\def\mom{{\cal H}^{(m)}}
\def\qham{\hat{\mathcal H}^{(m)}} %
\def\qmom{\hat{\mathcal{H}}^{(m)}}
\def\qhamint{\hat{H}^{(m)}}
\def\gravham{H^{(gr)}}
\def\matterham{H^{(m)}}
\def\qmatterham{\hat{H}^{(m)}}
\def\lapsh{[N,\vec{N}]}
\newcommand{\ann}[1]{{\hat{a}}_{\vec{#1}}}
\newcommand{\adag}[1]{{\hat{a}}^\dagger_{\vec{#1}}}

\def\0mom{{\rate^{\alpha\beta}(\z)}}
\def\1mom{{\rate^{\alpha\beta}_1(\z)}}
\def\2mom{{\rate^{\alpha\beta}_2(\z)}}
\def\cqadm{{\Hq_{ADM}}}

\def\cqhamcon{{\mathcal L}}
\def\cqmomcon{{\mathcal L}_{a}}

\def\m{\mu}
\def\dt{\delta t}

\def\Hc{H}
\def\q{{q}}
\def\p{{p}}
\def\dz{{d\z}}
\def\rate{{W}}
\def\linrate{{\lambda}}
\def\psstay{{\gamma}}
\def\pshsgo{{\gamma}}
\def\ab{^{\alpha\beta}}

\newcommand{\M}[1]{M_{#1}\ab}

\def\lax{{\L_\alpha(x)}}
\def\lbx{{\L^\dagger_\beta(x)}}
\def\lanox{{\L_\alpha\xd}}
\def\lbnox{{\L^\dagger_\beta\xd}}
\def\lbxp{{\L^\dagger_\beta(x')}}
\def\lby{{\L^\dagger_\beta(y)}}

\def\rateab{{\rate\ab(\z|\z';t)}}
\def\rateabx{{\rate\ab(\z|\z';x,y)}}

\def\rateabxd{{\rate_h\ab(\z|\z-\dist;x,y)}}
\def\hab{{h\ab(\z)}}
\def\habx{{h\ab(\z;x)}}

\renewcommand{\varrho}{\hat{\rho}}
\def\cqstate{\varrho}
\def\psiz{{\varrho(\z;t)}}
\def\psizt{{\varrho(\z;t)}}

\def\psizp{{\varrho(\z';t)}}
\def\psizd{{\varrho(\z-\dist;t)}}
\def\Lorentz{M}
\def\down{\downarrow}
\def\up{\uparrow}

\def\gnox{(g)}

\def\gpi{(g,\pi)}
\def\gpit{(g,\pi;t)}

\def\D{\nabla}

\def\PB{\}}
\def\PBg{\}_g}
\def\PBm{\}_m}
\def\friction{\gamma}
\def\hatA{\hat{A}}
\def\t0{0}
\def\bfa{\hat{a}}

\def\xd{}%
\def\justifying{%
	\rightskip=0pt
	\spaceskip=0pt
	\xspaceskip=0pt
	\relax
}

\AtBeginDocument{\renewcommand{\natexlab}[1]{#1}}
\begin{document}

\title{A postquantum theory of classical gravity?}

\author{Jonathan Oppenheim}
\affiliation{Department of Physics and Astronomy, University College London, Gower Street, London WC1E 6BT, United Kingdom}

\begin{abstract}
	The effort to discover a quantum theory of gravity is motivated by the need to reconcile the incompatibility between quantum theory and general relativity. Here, we present an alternative approach by constructing a consistent theory of classical gravity coupled to quantum field theory. The dynamics is linear in the density matrix, completely positive and trace preserving, and reduces to Einstein's theory of general relativity in the classical limit.  Consequently, the dynamics doesn't  suffer from the pathologies of the semiclassical theory based on expectation values. The assumption that general relativity is classical necessarily modifies the dynamical laws of quantum mechanics -- the theory must be fundamentally stochastic in both the metric degrees of freedom and in the quantum matter fields. This allows it to evade several no-go theorems purporting to forbid classical-quantum interactions. The measurement postulate of quantum mechanics is not needed -- the interaction of the quantum degrees of freedom with classical space-time necessarily causes decoherence in the quantum system. We first derive the
general form of classical-quantum dynamics and consider realisations which have as its limit deterministic classical Hamiltonian evolution. The formalism is then applied to quantum field theory interacting with the classical space-time metric. One can view the classical-quantum theory as fundamental or as an effective theory useful for computing the back-reaction of quantum fields on geometry. We discuss a number of open questions from the perspective of both viewpoints. 
\end{abstract}
\maketitle

\section{Must we quantise gravity?}

The two pillars of modern theoretical physics are general relativity, which holds that gravity is the bending of space-time by matter, and quantum field theory, which describes the matter that live in that space-time. Yet they are fundamentally inconsistent. Einstein's equations for gravity
\begin{align}
G^{\mu\nu}
=\frac{8\pi G}{c^4}{T}^{\mu\nu}
\label{eq:Es}
\end{align}
has a left-hand side which encodes the space-time degrees of freedom via the Einstein tensor and is treated classically, while on the right-hand side sits the energy-momentum tensor encoding the matter degrees of freedom which must become an operator $\hat{T}^{\mu\nu}$ according to quantum theory (we will henceforth denote quantum operators by placing a $\hat{}$ on them)\phantomsection\label{par:hat}.
The widespread belief is that this inconsistency should be remedied by finding a quantum theory of gravity so that the left-hand side of Equation \eqref{eq:Es} is also an operator. Yet, although we have candidates such as string theory, which is in its mid-50's\cite{veneziano1968construction}, and loop quantum gravity just over 40\cite{sen1982gravity,ashtekar1986new,rovelli1988knot}, a convincing theory of quantum gravity remains elusive.

Another possibility is that gravity should still be treated classically, but that Einstein's  equation should be modified. There is some sense to this. Space-time, though dynamical, can be understood as describing the background on which quantum fields of matter interact. Gravity, through the equivalence principle, is unique in this regard --
no other fields can be described as a universal geometry in which quantum fields live.  
Having this classical background structure appears crucial to our current understanding of quantum mechanics. Quantum field theory allows us to predict the result of future measurements on a space-like surface, based on past initial data. The identification of a Cauchy surface on which to give this initial data, not to mention the identification of a one parameter family of future hypersurfaces, is natural when the background structure is classical, while it is unclear whether it is possible to quantise this causal structure in a way which is independent of these background choices. While this difficulty may be merely technical, background independent approaches such as loop quantum gravity, {\it String Field Theory}\cite{PhysRevD.46.5467}, and others\cite{hardy2005probability,oreshkov2012quantum} have yet to demonstrate that this is possible.
And while the ultraviolet 
catastrophe required the quantisation of electromagnetism, the analogous issue in gravity is less well understood\cite{smolin1984thermodynamics} and muted by the fact that gravitational systems in equilibrium occur in exotic situations\cite{hawking1983thermodynamics}, or would emit gravitational radiation over very long time scales\cite{Simidzija_2021}. \phantomsection\label{par:uv}

Nonetheless, there is no strong case that gravity is exceptional, and the overwhelming consensus has been that we must quantise it along with all the other fields. In part, this is due to 
a number of no-go theorems and arguments over the years\cite{bohr1933on,cecile2011role,dewitt1962definition,
eppley1977necessity,caro1999impediments,salcedo1996absence,
sahoo2004mixing,terno2006inconsistency,salcedo2012statistical,barcelo2012hybrid,marletto2017we} purporting to require the quantisation of the gravitational field.
A recent no-go theorem\cite{marletto2017we} holds not only for coupling classical mechanics with quantum theory, but also with modifications to quantum-mechanics, so called post-quantum theories which alter the quantum state space or dynamics\cite{dynamics_foot}.
But given the difficulties faced in constructing such a theory, it seems important to revisit this consensus. The question of why space-time might be fundamentally classical in contrast to other forces will be considered elsewhere\cite{oppenheim2023rethink}, while here we merely consider the necessity of quantising gravity. Indeed, not only will we find that it is it logically possible to have quantum fields interacting with classical space-time, we will construct such a theory. The no-go theorems on the need to quantise gravity implicitly or explicitly assume the coupling of quantum fields to space-time is not stochastic. This loophole can be exploited to construct a theory of classical space-time which can be acted upon by quantum fields.

\begin{figure}[h]
\includegraphics[width=0.8\textwidth]{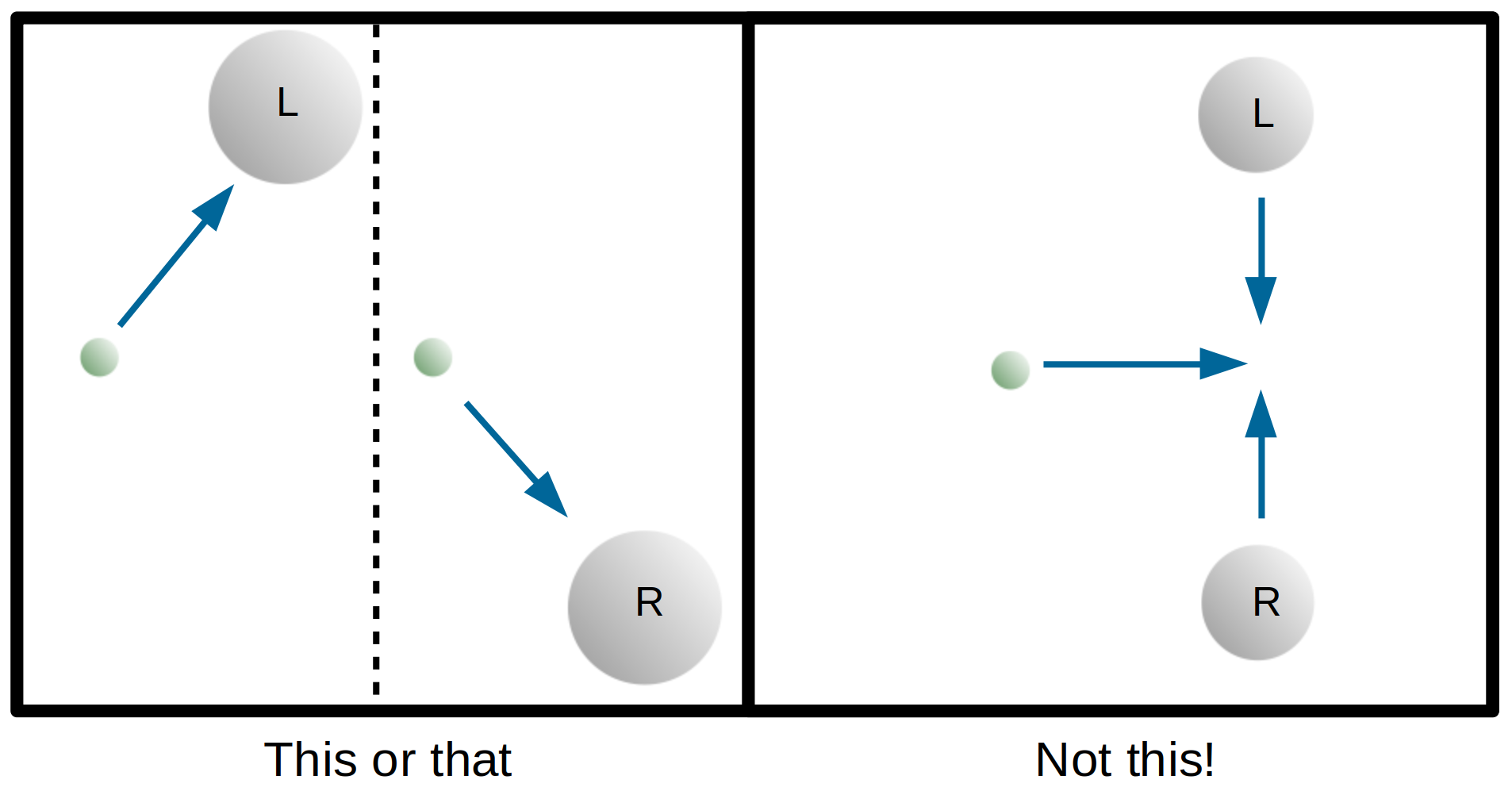}
\caption{The prediction of the semiclassical Einstein's equation is depicted in the right panel, compared with what we expect to occur based on the statistical interpretation of the density matrix in the left panel. If a planet is in a statistical mixture of being in two possible locations (``L'' and ``R'') then we expect the gravitational field to become correlated with it. A test particle therefore falls toward one of the planets. If one treats the semiclassical Einstein's equations as fundamental, the test particle falls down the middle which is indeed the average trajectory of its path in the left-hand panel. The planets are also attracted toward the place they might have been. The systems in the linear theory considered here exhibit the behaviour depicted in the left-hand panel.
The case where we attempt to put the planet in a superposition of the two locations is discussed further in
\cite{layton2022semi,UCLcoherence}. If the gravitational field is suitably different in the two cases, in the sense of how distinguishable the field configurations  are when noise is added to them, then there will be a limit on how coherent the superposition can be\cite{UCLcoherence}. In the case where there is coherent superposition, the initial trajectory of the test mass will be insufficiently correlated with the planet's position to enable a determination of where the planet is. However, after the decoherence time, it may be possible to determine the planet's location from the trajectory of the test mass, since it's trajectory will become correlated to the planet's location\cite{layton2022semi}. This is also what one expects from a fully quantum theory, with the stochasticity of the test-mass being due to vacuum fluctuations of the metric. %
}
\label{fig:notthis}
\end{figure}

Although the construction of a consistent fundamental theory is the principal motivation of the present work, two other motivations remain even if one believes that the fundamental theory of gravity is quantum. The first is to better understand the back-reaction that quantum fields have on space time. In many cases, such as in the evaporation of large black-holes, or in inflationary cosmology, we are interested in the limit in which space-time can be treated classically, while the matter itself, such as the black-hole radiation, or vacuum fluctuations, must be treated quantumly. We would like to compute the back-reaction that quantum matter produces on space-time, yet we have no consistent theory which enables us to do this. The current approach to describe gravity classically is to take the expectation value of the right-hand side of Equation \eqref{eq:Es}, leading to the  semiclassical Einstein's %
equations\cite{sato1950attempt,moller1962energy,rosenfeld1963quantization}
\begin{align}
G^{\mu\nu}
=\frac{8\pi G}{c^4}\langle{\hat T}^{\mu\nu}\rangle
\label{eq:semi}
\end{align}
Although this equation is often used to understand the back-reaction on space-time, it is only valid when fluctuations are small, while to understand quantum effects, we are precisely interested in the case where quantum fluctuations are significant. 
Taking the semiclassical Einstein's equation seriously leads to pathological behaviour. If a heavy object is in a statistical mixture of being in two locations (as depicted in Figure \ref{fig:notthis}), a test mass will fall toward the point between the two locations and the object itself will be attracted to the place where it might have been. This has been ruled out experimentally, in one of the most sublime  examples of trolling ever published in Physical Review Letters\cite{page1981indirect}. %

This problematic behaviour of the semiclassical Einstein's equation has generally been viewed as a reason why we must quantise gravity\cite{dewitt1953new,duff1980inconsistency,unruh1984steps,carlip2008quantum}. However, one sees the same apparent paradox in a fully quantum or fully classical theory of two systems. Take for example the interaction Hamiltonian $\hat{H}_{MG}=\hat{H}_M\otimes \hat{H}_G$, and the limit where the $M$ and $G$ system are close to a product state with density matrix $\hat{\rho}_{MG}\approx\hat{\sigma}_M\otimes\hat{\sigma}_G$.
Then for as long as the two systems remain approximately uncorrelated, Heisenberg's equation of motion for the $M$ system, after tracing out the $G$ system gives
\begin{align}
\frac{\partial \hat{\sigma}_M}{\partial t}\approx -i[\hat{\sigma}_M,\hat{H}_M]\otimes\langle \hat{H}_G \rangle
\label{eq:homer-semi}
\end{align}
where we use units where $\hbar=c=16\pi G=1$.
The state of the system $M$ also appears to be coupled to an expectation value, even though we know that this is not what happens. By tracing out the $G$ system, we have lost view of the fact that the two systems will get correlated. Once the systems become correlated, the reduced dynamics is no longer even linear in the reduced state $\hat{\sigma}_M$ since $\tr_G \hat{H}_G\hat{\rho}_{MG}\neq \langle \hat{H}_G\rangle\hat{\sigma}_M$. 
While this in no way suggests that the semiclassical Einstein's equation are correct even on average, its pathologies have led to a widely held objection to theories which treat gravity classically, while it is clear from the above example that the pathologies of the semiclassical Einstein's equations have nothing to do with treating one system classically, and everything to do with taking expectation values, and thus losing sight of correlations. The same apparent pathologies occur if both systems are classical but we allow for probabilistic mixtures of states. The theory presented here doesn't suffer from these issues -- the equations of motion are linear in the density matrix, just like quantum mechanics and the classical Liouville equation.  In order to understand the interaction of two systems, whether they are both quantum, both classical, or are classical-quantum, we need to account for the correlations between them, and this is more easily done by considering the evolution of a joint density matrix or probability density.

This leads us to the second motivation for the present work: even if one doesn't regard the theory as fundamental it provides a sandbox, or toy-model for understanding quantum gravity.  Indeed, a natural first step to quantising gravity is to begin with the Liouville equation for general relativity, and a probability density $\rho(t)$ over its phase-space.  After all, the probability density and the Liouville equation have a lot in common with the quantum density matrix and the Heisenberg equations of motion. As we have just seen with the semiclassical Einstein equation, some of the conceptual difficulties faced in understanding quantum gravity are also found in classical probabilistic theories and are resolved by considering instead the Liouville equation. The classical limit of the theory we present here, should thus be close to the Liouville equation for classical Einstein gravity, introduced as Equation \eqref{eq:GRLiouville}.
Many issues which arise in quantum gravity, especially in the canonical quantisation framework, will turn out to have a similar but more tractable form here. In particular, the difficulty encountered in understanding the role of diffeomorphism invariance when one doesn't have a single 4-geometry or single time-slicing, as well as the technical difficulty encountered in calculating the constraint algebra, also appear in the theory considered here. And even in some cases, in the fully classical limit of the theory.

\begin{figure}[h]
\includegraphics[width=0.8\textwidth]{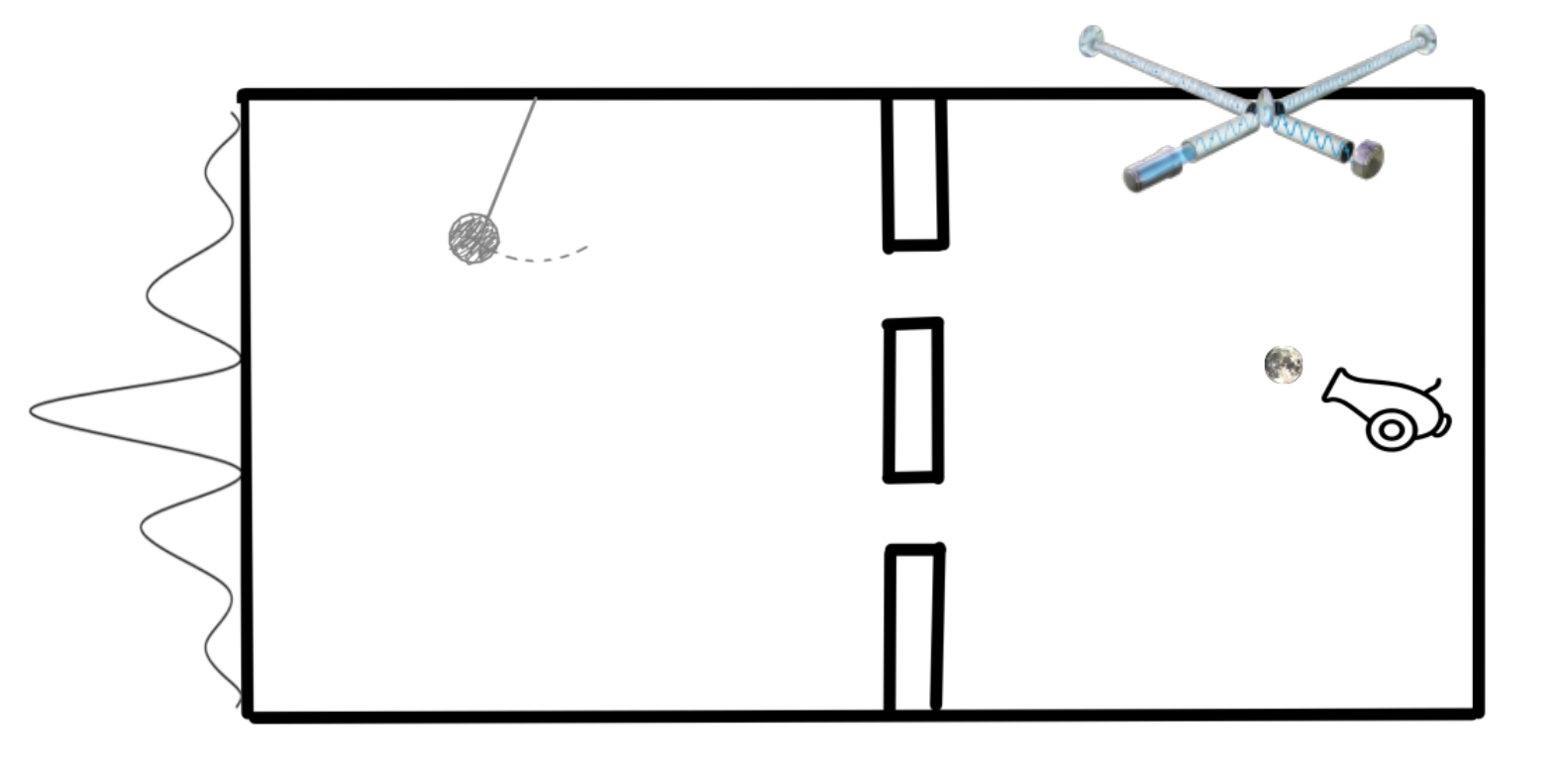}
\caption{Consider a variation of the {\it gedanken experiment} proposed by Feynman\cite{cecile2011role,Feynman:1996kb-note} and Aharonov\cite{AharonovParadoxes-note} in which a massive particle travels through two slits as a plane wave. Quantum theory predicts an interference pattern on the screen behind the slits which enables a determination of the particle's momentum. 
The local Newtonian potential it creates can be measured everywhere. 
	At late times, the state of the electromagnetic field far from the slits will be slightly different depending on which slit the particle goes through, but these two states, need not be orthogonal. The quantum nature of the electromagnetic field prevents an experimenter from distinguishing these two non-orthogonal states and while the electromagnetic interaction might lead to some decoherence of the particle, if the interaction is small enough the interference pattern remains. On the other hand, if the gravitational field is classical and the interaction deterministic, then one could imagine measuring the gravitational field close to the particle and during its flight, without disturbing the field to unambiguously determine which path the particle went through. One could therefore never get interference fringes (or the uncertainty principle would be violated). Two classical states cannot be different but non-orthogonal. Actually measuring the gravitational field is not required since classical systems are undisturbed -- even if we can't actually measure the field to the required accuracy\cite{mattingly2006eppley}, it is impossible to write down a pure quantum state of the particle which is correlated with a classical gravitational field determined by the particle's position. Since the theory presented here modifies quantum theory so that the interaction is stochastic, the gravitational field need not become unambiguously correlated with the particle's path and the interference pattern is restored. Probability densities over classical states can be different and non-orthogonal.}
\label{fig:doubleslit}
\end{figure}

Let us now turn to another widely held belief on why a classical theory of gravity is incompatible with superpositions and the uncertainty relation. This argument is usually attributed to Feynman\cite{cecile2011role,Feynman:1996kb-note} and Aharonov\cite{AharonovParadoxes-note} in relation to the gravitational field (see also \cite{eppley1977necessity,Mari:2015qva,Baym3035,belenchia2018quantum,signal_foot}). As depicted in Figure \ref{fig:doubleslit}, if the gravitational field produced by a particle is classical, the field can in principle be measured to sufficiently high precision  
to determine the particle's position without disturbance. This would reveal which slit it went through, and would either prevent particles from being in superposition or would violate the uncertainty principle if the interference experiment of Figure \ref{fig:doubleslit} is carried out.  However, if the classical field responds to the quantum system stochastically\label{par:stochastic}\cite{hustochasticgravity_foot}, then measuring the classical degrees of freedom will not necessarily determine the particle's quantum state to arbitrary precision. This could be because the gravitational field reacts in an indeterministic way to the presence of matter,  or because there is only a probability that the gravitational field reacts to the position of the particle during the time of the interference experiment\cite{experiment_foot}. In either case, the gravitational degrees of freedom only contain partial information about the location of the particle. We thus see that previous arguments for requiring the quantisation of the space-time metric implicitly assume that theory is deterministic, and are not a barrier to the theory considered here.

The fact that coupling classical gravity to quantum theory necessarily requires stochasticity\cite{stochasticity_foot} 
 is particularly compelling in light of the black-hole information problem\cite{hawking-bhinfoloss,hawking-unpredictability,preskill-infoloss-note}, and its sharper version, the AMPS ''paradox'' \cite{almheiri2013black,braunstein2009entangled}. Since deterministic theories appear to require a breakdown of gravity at low energies\cite{energies_foot}, serious consideration to non-deterministic theories should be given\cite{ellis1984search}. However theories with information destruction face imposing obstacles. \phantomsection\label{par:introchallenge} Banks, Peskin and Susskind (BPS) argued that they lead either to violations of locality or anomalous heating
\cite{bps,gross1984quantum}, and Coleman argued that they merely result in false decoherence corresponding to unknown coupling constants\cite{coleman1988black}. 
\phantomsection\label{par:ibps}There have been a few attempts to work around the issue raised by BPS. 
Unruh and Wald proposed that if the
information destruction was at sufficiently high energy, then violations in momentum and energy conservation 
would also be at high energy, and hence, not observable in the lab\cite{unruh-wald-onbps}. Models which are local, yet violate 
cluster-decomposition\cite{cluster_foot}
were proposed in \cite{OR-intrinsic}, and ones which have only mild violations of energy conservation at the expense of some non-locality were proposed in \cite{poulinKITP}. 
However, none of these attempts have been theories of gravity, and once the back-reaction to space-time is considered and the constraints of general relativity taken into account, the situation is different enough to warrant reconsidering the issue. The class of theories presented here may allow us to resolve the black-hole information paradox in favour of information loss. We shall return to these points in the Discussion section.

We will find that the issue of anomalous heating takes on a different form in this theory,
	but continues to raise questions. These may be resolved simply by regularising the theory, but it is also possible that we will need to introduce new physics at a distance scale larger than the Planck scale, as needs to happen in spontaneous collapse models\cite{ballentine1991failure,gallis1991comparison,shimony1990desiderata,donadi2021underground}. %
	Further  research is needed, because the theory we present will contain a number of unspecified couplings and choices. These can be shown to be greatly reduced if one demands that the theory should be continuous in phase-space\cite{UCLPawula}. However, this does not completely fix the theory, and there remains some additional freedom which we expect to be constrained by demanding the theory be regularisable, and by experimental constraints. Whether there is sufficient freedom in the couplings to satisfy these demands, as well as ones coming from symmetry principles such as diffeomorphism invariance, remains open.

Having argued that nothing forbids the coupling of quantum systems to classical ones, let us turn to attempts to construct such a theory.
An early attempt to couple toy models for classical gravity to quantum mechanics was made in \cite{boucher1988semiclassical}
using the Aleksandrov-Gerasimenko (AG) bracket\cite{aleksandrov1981statistical,gerasimenko1982dynamical} 
\begin{align}
\frac{\partial \psiz}{\partial t}
=
-i[\Hq(\z),\psiz]+\frac{1}{2}\Big(\{\Hq(\z),\psiz\PB-\{\psiz,\Hq(\z)\PB\Big)
\label{eq:alex}
\end{align}
where $\hat{H}$ is a quantum Hamiltonian which can depend on phase space points $\z$, and $\{\cdot,\cdot\}$ are Poisson brackets. The object $\psiz$, about which we will say more in Section \ref{sec:cq-review}, is the classical-quantum (CQ) state at time $t$ -- a positive matrix in Hilbert space, and a functional of the classical degrees of freedom $\z$, normalised such that
\begin{align}
\int d\z \tr\psiz=1\,.
\label{eq:normalisation}
\end{align}
However, 
the dynamics the AG bracket generates leads to negative probabilities\cite{boucher1988semiclassical,diosi2000quantum}. Unless we are prepared to modify the Born rule, or equivalently, our interpretation of the density matrix, any dynamics must be trace-preserving (TP) to conserve probability, and be {\it completely positive (CP)}, for probabilities to remain positive. So, although the dynamics of Equation \eqref{eq:alex} makes a frequent appearance in attempts to couple quantum and classical degrees of freedom 
\cite{anderson1995quantum,kapral1999mixed,kapral2006progress} (c.f. \cite{prezhdo1997mixing}), and while it may give insight in some regimes, it cannot serve as a fundamental theory. 
Other attempts, such as using the Schr\"odinger-Newton equation\cite{diosi1987universal,penrose1998quantum} (c.f. \cite{pitaevskii1961vortex,gross1961structure}), suffer from the problem that the equations are non-linear in the density matrix, and thus will lead to superluminal signalling\cite{gisin1989stochastic,gisin1990weinberg,polchinski1991weinberg} and a breakdown of the statistical mechanical interpretation of the density matrix\cite{matrix_foot}.\label{par:antag2}

And yet experimentalists working in the field of quantum control routinely couple quantum and classical degrees of freedom together, usually by invoking the mysterious measurement postulate of quantum theory. This presents another path to treat gravity classically, where one 
imagines the system of interest is fundamentally quantum, but that observables are restricted or a measurement is performed on them
\cite{jauch1964problem,diosi1998coupling,diosi2000quantum,peres2001hybrid}. However, for gravity, this would require first quantising the theory, which might well be an impossible task. Another approach, is to apply a measurement to the quantum system instead, and then couple the output to the classical degrees of freedom such as the Newtonian potential\cite{kafri2014classical,tilloy2016sourcing,tilloy2017principle,dowker2008dynamical}.  Both these approaches, at least initially, invoked the measurement postulate of quantum mechanics, or an ad-hoc field which produces a stochastic collapse of the wave function
\cite{pearle-csl1,grw85,grw86,gisin-percival,adler2001generalized,sep-qm-collapse} which then source the gravitational field. 

And here lies the essence of the problem -- if nature is fundamentally quantum, our notion of classicality hinges on the measurement postulate. Thus the emphasis in this work, is to invert the logic: assume that since space-time describes causal structure and relationships between the matter degrees of freedom, it is a-priori and fundamentally classical. Then examine the implications of this. One outcome will be that the quantum degrees of freedom inherit classicality from space-time. By treating the space-time metric as a classical degree of freedom, we obtain for free not only a 
theory of quantum matter and gravity, but also a theory which produces the
 gravitationally induced collapse of the wave-function, conjectured by Karolyhazy, Diosi and Penrose\cite{karolyhazy1966gravitation,diosi1989models,penrose1996gravity}\cite{penrose_foot}. Collapse arises as a simple consequence of treating space-time classically. Such an approach can be found in \cite{hall2005interacting} where a variation of Equation \eqref{eq:semi} is studied, but it is argued that the interaction of a quantum system with a classical measuring device necessarily leads to decoherence and consistency of the measurement postulate. Likewise Diosi and Tilloy\cite{tilloy2016sourcing,tilloy2017principle} point out that although they introduce a spontaneous collapse model to source the Newtonian potential, this can be thought of as formal, and it is the consistency of hybrid dynamics which requires decoherence of the quantum matter.  Poulin \cite{poulinKITP} also notes that the classical coupling leads to decoherence of the quantum system.
 
The theory considered here however, cannot be reduced to a linear spontaneous collapse model, because the dynamics is not Markovian or even linear in the density matrix when the gravitational degrees of freedom are traced out (although it is linear and Markovian when we include the metric degrees of freedom). It also leads not merely to decoherence, but
to a genuine objective collapse of the wave-function.
In particular, there are realisations of the discrete theory considered here, where stochastic jumps in phase-space can be finite and correlate to Lindblad operators. As a result, changes in the quantum state correspond to changes in the classical system. As a a result, the quantum state remains pure and uniquely determined by the trajectory of the classical system~\cite{UCLqubit}. Purity of the quantum system conditional on the classical trajectory has since been shown to also hold in the continuous realisations under natural conditions\cite{layton2022semi}. 

Before proposing a theory which couples classical general relativity with quantum field theory, we first construct a
general framework for classical-quantum dynamics.
 In order to do so, we will derive general linear dynamics which maps the set of classical-quantum states to itself. More precisely, we wish to find the most general linear dynamics which is completely positive, trace-preserving and preserves the division of degrees of freedom into classical and quantum. Two examples of such dynamics have previously been proposed. In \cite{blanchard1993interaction,blanchard1995event} a master equation with diagonal coupling to Lindblad operators $\L_\alpha$  was considered, of the form
\begin{align}
  \frac{\partial\psiz}{\partial t}
  =&-i[\Hq(\z),\psiz]+
  \sum_{\alpha,\z'}  n
 \rate^{\alpha}(\z|\z')
  \L_{\alpha}\psizp\L_{\alpha}^\dagger  
-\frac{1}{2}\rate^\alpha(\z)\{\L_\alpha^\dagger\L_\alpha,\psiz\}_+
      \label{eq:BJAKDP-dynamics}
\end{align} where $\z$ are discrete classical degrees of freedom,
 $\{\cdot\}_+$ is the anti-commutator. 
When $\rate^\alpha(\z|\z')=W^\alpha(\z)\delta(\z-\z')$, this reduces to
the GKSL or Lindblad equation\cite{GKS76,Lindblad76} controlled by an external degree of freedom $\z$.  Equation \eqref{eq:BJAKDP-dynamics} has been shown by Poulin to be the most general master equation in the case of discrete classical variables and bounded Lindblad operators by embedding the classical system in Hilbert space\cite{poulinPC}. However, for trajectories which are continuous in the classical degrees of freedom, it will turn out that non-diagonal couplings are required. A master equation due to Diosi corresponds to the constant force case\cite{diosi1995quantum}. 
This can be generalised to the continuous master equation since derived in \cite{UCLPawula}
\begin{align}
\frac{\partial\psiz}{\partial t}
=&\{H_0(\z),\psiz\}-i[\Hq(q),\psiz]
+
\frac{1}{2}\Big(\{\Hq(q),\psiz\PB-\{\psiz,\Hq(q)\PB\Big)
 \nonumber\\ 
&+\frac{1}{2}\int dxdx' \{q_i(x'),\{q_j(x),D_2^{ij}(x,x')\psiz\}\}
+\frac{1}{2}\int dxdx'D_{0,ij}(x,x')
\left[\{\Hq(q),p^i(x)\},\left[\psiz,\{\Hq(q),p^j(x')\}\right]\right]
\label{eq:CQcontinuousHam}
\end{align}
which adds diffusion and a Lindbladian to the Aleksandrov-Gerasimenko bracket of Equation \eqref{eq:alex}, and is completely positive provided the matrix inequality $2D_2\succeq D_0$ is satisfied and $(\mathbb{I}- D_0 D_0^{-1}) =0$.  For convenience, the classical-quantum interaction Hamiltonian $\Hq(q)$ is chosen to be minimally coupled, meaning that it is a functional of phase space fields $q_i(x)$ and not on the conjugate momenta $p_i(x)$, while $H_0(\z)$ is any purely classical Hamiltonian.
The constant force case considered in \cite{diosi1995quantum} corresponds to an interaction Hamiltonian which is linear in a single classical degree of freedom $q(x)$,  $\Hq(q)=\int dx q(x)\L(x)$.  The constant force case has
been studied in the context of measurement theory\cite{diosi2014hybrid}, in the classical limit of the collisionless Boltzmann equation\cite{alicki2003completely}, and in a model for Newtonian gravity\cite{diosi2011gravity}. The discrete dynamics has appeared in discussions around black-hole evaporation\cite{poulinKITP}. %

 In Section \ref{sec:cq-dynamics}, we show that Equation \eqref{eq:cq-dynamics} is the most general form of  {\it time-local}\cite{ktimelocal_foot}
  classical-quantum dynamics. It includes the discreet dynamics of Equation \eqref{eq:BJAKDP-dynamics} and the continuous Hamiltonian dynamics of Equation \eqref{eq:CQcontinuousHam}.
 Our derivation uses a fully classical-quantum formalism and gives necessary conditions for the dynamics to be completely positive. %
These are also sufficient in the case of discrete dynamics and in \cite{UCLPawula} with  Šoda, Sparaciari and Weller-Davies we derive the necessary and sufficient conditions when the dynamics is continuous in phase space.\label{par:necsuf} This continuous master equation generalises the master equation of \cite{diosi1995quantum} to the case of arbitrary interaction Hamiltonians and multiple Lindblad operators and classical degrees of freedom. Since the continuous master equation requires off-diagonal Lindblad couplings, this can be shown to imply that the dynamics of Equation \eqref{eq:BJAKDP-dynamics}  has discrete, finite sized jumps in phase space\cite{UCLPawula}. 

With the general form of dynamics in-hand, the next step will be to restrict ourselves to dynamics which is generated by a Hamiltonian. This will allow us to apply the framework to the Hamiltonian formalism of general relativity. 
This follows from considering a perturbative expansion of the more general dynamics, presented as Equation \eqref{eq:weiner2}.  This expansion can be thought of as a CQ version of the Kramers-Moyal expansion\cite{kramers1940brownian,moyal1949stochastic}, familiar in classical stochastic dynamics.
We show in Section \ref{sec:Hdynamics}, that continuous and deterministic classical evolution emerges in some limit. 
Away from this limit, the quantum system can retain some coherence, but complete positivity requires the classical system to undergo diffusion or discrete, finite sized jumps in phase-space. 
A discrete realisation of this dynamics makes use of a fininte and non-commuting directional divergence, Equations \eqref{eq:PQdiv-simple}-\eqref{eq:non-commuting_divergence-lin}. This generates 
a discrete version of Liouville dynamics on the classical system and the action of a dynamical semi-group on the quantum system, thus providing a natural generalisation of both classical and quantum evolution. 
A parameter $\tau$ governs how continuous and classical the total system is, with the system's trajectory in the classical phase-space going from finite-sized hops to continuous deterministic dynamics as $\tau\rightarrow 0$. In this limit, the quantum system decoheres and becomes classical. For larger $\tau$ (or $\hbar$), the decoherence times are longer, while the jump distance in the classical phase-space increases resulting in greater diffusion in the system's trajectory. The dynamics of Equation \eqref{eq:weiner2} also includes the fully continuous evolution of Equation \eqref{eq:CQcontinuousHam}.

In Section \ref{sec:qubit}, we present a simple example of the discrete dynamics, namely a quantum spin system with a classical position and momentum interacting with a potential. We see that if the particle is initially in a superposition of spin states, then it eventually collapses to one of them through its interaction with the classical degrees of freedom. In essence, the classical potential acts like  a Stern-Gerlach device, so that after some time, the particle's classical degrees of freedom uniquely determine the value of the spin. Unlike standard decoherence, the collapse of the quantum state happens suddenly, when the system undergoes a classical jump in its momentum. At that point, monitoring the momentum unambiguously reveals the particle's spin. The quantum state of the system remains pure, conditioned on the state of the classical system. There is thus no need for the mysterious measurement postulate of quantum theory, a postulate which has meant that the interpretation of quantum theory has been open to dispute, and whose problematic nature has  received renewed attention\cite{frauchiger2016single}.

In Section \ref{ss:PQFT} we then extend the framework to include a hybrid version of quantum field theory. In Section \ref{sec:PQgrav} we apply this hybrid formalism to the case of general relativity. We work in the ADM formalism\cite{arnowitt2008republication}, treating the metric as classical, while the matter living on space-time are quantum fields. %
	In the classical limit, the  equations of motion reduce to Einstein's equations, in the sense of a CQ version of Ehrenenfest's theorem, that after taking the expectation value on the quantum fields, the equations of motion are close to the semiclassical version of the Liouville Equation (Equation \eqref{eq:semiclassical-mastereqn}), applied to general relativity (Equation \eqref{eq:GRLiouville}).
There will nonetheless be some deviations from general relativity in the classical limit, because there are additional contribtions relating to diffusion and decoherence, with some of the experimental bounds on these terms since given in \cite{oppenheim2021gravitationally}.
\phantomsection\label{par:Climit}

The equations we derive allow one to calculate the back-reaction of quantum fields on classical space time. This allows us to go beyond the semiclassical Einstein's equation and we expect it to prove useful in attempts to understand black-hole evaporation or the effect of vacuum fluctuations in cosmology.
 The interaction of the quantum system with classical space-time causes the wave function to collapse, at a rate determined by the system's energy momentum tensor. The stochastic nature of the interaction allows for the possibility that information is destroyed in black-holes, although the quantum state conditioned on the metric degrees of freedom can remain pure. The theory suggests a number of possible experiments which could falsify or verify it, and we discuss this in Section \ref{sec:discussion}.
We now begin in Section \ref{sec:cq-review} by reviewing the formalism of classical-quantum dynamics and then discuss how the evolution law we will derive relates to both classical stochastic dynamics and quantum maps. The form of the master equation, Equation \eqref{eq:cq-dynamics}, is then derived in Section \ref{sec:cq-dynamics}. 
Since the initial posting of this manuscript in 2018\cite{oppenheim2018postArxiv}, a number of advancements have been made. We will continue to highlight some of these developments while maintaining the integrity of the original draft.

\section{Consistent classical-quantum dynamics}
\label{sec:cq-review}

The state of a quantum system is represented in a Hilbert space $\mathcal{H_S}$ by a density matrix -- a positive trace one matrix $\hat{\sigma}(t)$ at time $t$.  A discrete %
classical system is represented by a discrete random variable $\z$, and the system has a probability $p(\z;t)$ of being at $\z$.  In the continuous case, we will consider a classical system to live in phase-space although we could just as well consider any other classical sample space. Phase-space is a 2n dimensional manifold ${\mathcal M}$ of points $\z=(\q,\p)^T$ and the state of the classical system can be represented by a probability density -- a nonnegative distribution %
 $\rho(\z;t)$ defined over $\z$ at time $t$, normalised such that $\int d\z \rho(\z;t)=1$.

Here, as in \cite{aleksandrov1981statistical,gerasimenko1982dynamical, blanchard1993interaction,blanchard1995event,diosi1995quantum, diosi2014hybrid,alicki2003completely,poulinKITP}, we consider a hybrid classical-quantum system which lives on both the Hilbert space and the classical phase space, an illustrative example being the classical-quantum qubit of Equation \eqref{eq:cqQubit}. The state of a hybrid system $\cqstate(t)$ at time $t$ is defined over an interval around $\z$ by a positive  matrix $\cqstate(\z;t)$ in $\mathcal{H_S}$\cite{barb_foot}, such that the normalisation condition Equation \eqref{eq:normalisation} holds. 
In quantum theory, such a system is sometimes said to be in a {\it cq-state} (or CQ state), and has a density matrix of the form %
\begin{align}
\varrho_{cq}=\int d\z\rho(\z;t)\proj{\z}\otimes\hat{\sigma}({\z;t})  
\label{eq:cq-state}
\end{align}
with $\hat{\sigma}(\z;t)$ a normalised density matrix and $\ket{\z}$ an orthonormal basis.
Since $\hat{q}$ and $\hat{p}$ don't commute in quantum theory, if we want to 
encode phase space degrees of freedom in a quantum system, we could consider two separate subsystems, one which encodes $q$ and one which encodes $p$.\label{ft:embed}
Here, we will not consider such an embedding, but rather describe the classical-quantum
 state as 
 \begin{align}
 \cqstate(\z;t)=\rho(\z;t)\hat{\sigma}(\z;t)
 \label{eq:cqRef3}
 \end{align}

with %
 $\rho(\z;t):=\tr\psiz$ a phase space density and $\int\tr\psiz d\z \leq 1$ when integrated over any interval.\phantomsection\label{par:subnormal} 
Conditioned on the classical system being in state $\z$, the quantum system is in the state $\hat{\sigma}(\z;t)$, the normalised state of the quantum system. In finite dimension, we can also write the CQ state as a matrix whose diagonal entries are phase space densities, as with the CQ qubit of Equation \eqref{eq:cqQubit}.
When we consider gravity,
each point $\z$ is a classical field (the 3-metric, and 3-momentum) over $\mathbb{R}^3$.

\begin{table}
	\begin{center}
		\begingroup
		\setlength{\tabcolsep}{10pt} %
		\renewcommand{\arraystretch}{1.4}
		\begin{tabular}{|l| l|l|l|} 
	\hline
			&  Map & Positivity condition & Norm condition \\  
			\hline 
			Classical
			&  
			$\rho(\z;t) = \int dz' P(\z|\z';t) \rho(\z',0)$ 
			&	
			$P(\z|\z';t) \geq 0,\, \forall \z, \z'$ 
			&
			$\int d\z P(\z|\z', t) =1  $ \\
			\hline
			Quantum  
			&   $\hat{\sigma}(t) =  \lambda^{\alpha   \beta}(t)\L_{\alpha} \hat{\sigma}(0) \L_{\beta}^{\dag} $
			&
			$\lambda(t) \succeq 0$ 
			&
			$ \lambda^{\alpha \beta}(t) \L_{\beta}^{\dag} \L_{\alpha} =\mathbb{I} $ \\ 
			\hline
			Classical-Quantum
			&
			$\cqstate(\z;t) = \int d\z'  \Lambda^{\alpha \beta}(\z|\z';t)\L_{\alpha} \cqstate(\z';0) \L_{\beta}^{\dag} $ 
			&
			$\Lambda(\z|\z';t) \succeq 0,\,\forall z,z'$ %
			&
			$	 \int d\z\Lambda^{\alpha \beta}(\z|\z';t) \L_{\beta}^{\dag} \L_{\alpha} =\mathbb{I} $ 
			\\ 
			\hline
		\end{tabular}
		\caption{\label{tab: CPmapTable} Summary of Sections \ref{sec:cq-review}-\ref{sec:cq-dynamics} of the finite time map acting on states for stochastic classical, quantum and classical-quantum dynamics.   We use Einstein summation notation, so an index which appears both raised and lowered should be summed over. 
			In the classical case, the probability density $\rho(\z;t)$  over phase space $\z$ is given in terms of the earlier probability density $\rho(\z;0)$ and the transition probabilities $P(\z|\z';t)$ which define a classical channel. In the quantum case, the necessary and sufficient condition for complete positivity is that $\lambda(t)$ as a matrix in $\alpha\beta$ is positive semidefinite (which we write as $\lambda(t)\succeq 0$). $\{\L_\alpha\}$ are a basis of operators. The classical-quantum postivity and norm conditions combine those of the purely classical and purely quantum dynamics, and generalises the Kraus decomposition theorem to the hybrid case in Equation \eqref{eq:cq-kraus2}. }
		\endgroup
		
	\end{center}
\end{table}

We now ask, what is the most general dynamical map which take any cq-state at time $t=\t0$ to another cq-state at a later time $t$. This will require us to adapt Kraus's representation theorem\cite{Kraus} to the classical-quantum case. 
Table I compares the classical-quantum map to the Kraus map for quantum systems and the stochastic matrix for classical systems. Table II compares the corresponding master equations which we now review.

Let us recall that for quantum systems,
a linear map $\mathcal{E}(\hat{\sigma})$  which is completely positive trace preserving (CPTP) at every time $t$ (we say it is {\it infinitesimally divisible}\cite{wolf2008dividing}) is always generated by the GKSL or Lindblad equation\cite{GKS76,Lindblad76}
\begin{align}
\frac{\partial\hat{\sigma}(t)}{\partial t}=-\frac{i}{\hbar}[\hat{H}(t) ,\hat{\sigma}(t)]-\lambda^{\ag\bg}(t)\big[\L_\alpha\hat{\sigma}(t)\L_\bg^\dagger-\frac{1}{2}\{\L_\bg^\dagger\L_\ag,\hat{\sigma}(t)\}_+\big]
\label{eq:lindblad}
\end{align}
with ${ \hat{H}(t)}$ the Hamiltonian, the set $\{\L_\ag$\} a traceless basis of operators orthogonal in the Hilbert-Schmidt norm, and $\lambda^{\ag\bg}(t)$ are elements of a positive semidefinite (PSD) matrix which we write as $\lambda\succeq 0$. Here, and throughout, we use Einstein summation notation when there are raised and lowered indices. One can always diagonalise $\lambda$, in which case $\L_\ag\hat{\sigma}\L_\ag^\dagger$ is sometimes called the {\it jump term} since one could interpret it as giving the rate at which
the system jumps from being in another quantum state into $\hat{\sigma}$, while 
$\frac{1}{2}\{\L_\ag^\dagger\L_\ag,\hat{\sigma}\}_+$ is sometimes called the {\it no event} term and acts to preserve the trace. If the map is non-Markovian (or more precisely, not infinitesimally divisible), but has an inverse, then it can still be represented by Equation \eqref{eq:lindblad}, with the positive semidefinite condition on the $\lambda^{\ag\bg}(t)$ lifted, provided the finite time map is still CP\cite{time-local}. In finite dimensional Hilbert spaces, CP maps have an inverse, but this condition can fail in infinite dimension\cite{buzek1998reconstruction,breuer1999stochastic}, in which case, it may not be possible to write the master equation in time-local form.

For a classical stochastic process on the other hand, an initial state at $t=\t0$, $\rho(\z;\t0)$ evolves to
\begin{align}
  \rho(\z;t)=
\int d\z  
  P(\z|\z';t)\rho(\z';\t0)
  \label{eq:c-non-markovian}
\end{align}
with $P(\z|\z';t)\geq 0$ 
and 
$\int d\z P(\z|\z';t)=1$. This simply ensures that any initial probability density
$\rho(\z;0)$ remains positive
and normalised. This is equivalent to $P(\z|\z';t)$ being the conditional probability that the state makes a transition to $\z$ at time $t$ given that it was initially at $\z'$. For a time-local process, the most general dynamics is given by the rate equation
\begin{align}
    \frac{d\rho(\z;t)}{dt}=\int d{\z'} \rate(\z|\z';t) \rho(\z';t) - \rate(\z;t)\rho(\z;t)
    \label{eq:rate-eqn}
  \end{align}
  where $\rate(\z|\z';t)$ are rates for the system to transition from state $\z'$ to state $\z$ (this can be thought of as the {\it jump term}). To conserve probability, we require 
  \begin{align}
  \rate(\z;t)=\int d{\z'} \rate(\z'|\z;t)
  \label{eq:rateeqnnorm}
  \end{align} 
  being the
  total rate for the system to transition away from state $\z$ to one of the states $\z'$.
  $W(\z;t) \rho(\z)$  is analogous to the {\it no event} term in the GKSL equation. 
 
 The dynamics can be discontinuous in $\z$, containing discrete finite sized {\it jumps} in which case
 it is necessary and sufficient that $\rate(\z|\z';t)\geq 0$, $\forall z\neq z'$ for the dynamics to be positive.
 When  
  $W(\z|\z';t)=-D_{1,i}(\z')\partial_{z_i'}\delta(\z-\z')+
  D_{2,ij}(\z')\partial_{z_i'}\partial_{z_j'}\delta(\z-\z')$, the dynamics is continuous in $\z$, and  positive iff
  $D_{2,ij}$ is a positive semidefinite matrix in $ij$ corresponding to {\it diffusion}. $D_{1_i}$ is called the drift term\phantomsection\label{par:drift} and includes Hamiltonian evolution as a special case. $\rate(\z;t)=0$ by integration by parts in Equation \eqref{eq:rateeqnnorm}. 
 This gives the Fokker-Planck equation or Forward Kolmogorov equation, while the general form of the dynamics
is sometimes referred to as the {\it differential Chapman-Kolmogorov equation\cite{gardiner2004handbook}.}

\begin{table}
	\begin{center}
		\begingroup
		\setlength{\tabcolsep}{10pt} %
		\renewcommand{\arraystretch}{1.4}
		\begin{tabular}{|c| c|}
			\hline
			& Classical \\
			\hline
			Master equation & 
			$
			\partial_t\rho(\z;t)
			=\int d \z^{\prime} W\left(\z | \z^{\prime}\right) \rho\left(\z^{\prime};t\right)- \int d\z' W\left(\z'|\z\right) \rho(\z;t)$ 
			\\ 
			\cline{0-0}
			Positivity (jumps) &  $ W(\z|\z') \geq 0,\, \forall \z\neq \z'$  \\ 
			\cline{0-0}
			Positivity (continuous) & $W(\z|\z')=-D_{1,i}(\z')\partial_{z_i'}\delta(\z-\z')+
			D_{2,ij}(\z')\partial_{z_i'}\partial_{z_j'}\delta(\z-\z'),$ $D_2\succeq 0$
			\\
			\hline\hline
			& Quantum \\
			\hline
			Master equation  &  $%
			\partial_t\hat{\sigma}
			=-i[H, \hat{\sigma}]+h^{\ag \bg} L_{\ag} \hat{\sigma} L_{\bg}^{\dagger}-\frac{1}{2}\left\{h^{\ag \bg} L_{\bg}^{\dagger} L_{\ag}, \hat{\sigma}\right\}_+$  \\ 
			\cline{0-0}
			Positivity condition & $h^{\ag \bg}$ a positive semidefinite (PSD) matrix in $\ag,\bg$ ($h\succeq 0$)\\ 
			\hline
			\hline
			&   Classical-quantum \\
			\hline
			Master equation &
			$%
			\partial_t\psiz=\int d z^{\prime} W^{\alpha \beta}\left(z | z^{\prime}\right) L_{\alpha} \cqstate\left(z^{\prime};t\right) L_{\beta}^{\dagger}
			-\frac{1}{2}\int d z^{\prime} W^{\alpha \beta}\left(z^{\prime} | z\right) \{L_{\beta}^{\dagger} L_{\alpha}, \cqstate(z;t)\}_+$ 
			\\
			\cline{0-0}
			Positivity (jumps) & $W^{\alpha\beta}$ PSD in $\alpha\beta$  $\forall z\neq z'$ , $W^{\ag\bg}$ PSD in $\ag\bg$ for $\z=z'$ \\ 
			\cline{0-0}
			Positivity (continuous) & Equation \eqref{eq:continuous_pos_condition}
			\\
			\hline
		\end{tabular}%
		\caption{\label{tab: MasterEquationTable} Summary of Sections \ref{sec:cq-review}-\ref{sec:cq-dynamics} of the master equations governing classical, quantum and classical-quantum dynamics, along with the positivity condition in the case of continuous dynamics, or discrete jumps in the classical degrees of freedom. In the classical case, continuous dynamics corresponds to the Fokker-Planck equation, specified by a drift term $D_1(\z)$ and a positive semidefinite diffusion matrix $D_2(\z)\succeq 0$ with elements $D_{2,ij}(\z)$. 
				One can always choose $\L_0=\id$, and the remaining $\L_\ag$ to be traceless\cite{UCLPawula}.
				The CQ master equation, is a natural generalization of the classical rate equation (or Fokker-Planck equation in the continuous case) and the Lindblad equation, and is here presented as Equation \eqref{eq:cq-dynamics} (in this table we have used the convention of included the $\lambda^{\ag\bg}$ in the $\rate^{\ag\bg}$ terms). In general the coupling constants can have an explicit time-dependence which we have dropped for ease of presentation.}
		\endgroup
	\end{center}
\end{table}

Here we present the most general classical-quantum dynamical map which is linear and completely positive. In section \ref{sec:cq-dynamics}\ we find the generators of this map, deriving the most general continuous time master equation for the dynamics. As in the proof of the generality of the GKSL equation, the proof presented here is valid for a separable Hilbert space, and countable set of bounded Lindblad operators $\L_\alpha$, or alternatively, when the map has an inverse at infinitesimal short times.
In the case of unbounded operators\cite{siemon2017unbounded}, one can cast the equation in the form we derive, but it's possible the dynamics are not unique. %

Under these assumptions, we will find that the most general master equation can be put into the form
\begin{align}
  \frac{\partial\psiz}{\partial t}
  =&-i[\Hq(\z;t),\psiz]+
 \linrate^{\ag\bg}(\z;t)\Big(
\L_{\ag}\psizt \L_{\bg}^\dagger
  -
  \frac{1}{2}\{\L_\beta^\dagger\L_\alpha,\psiz\}_+
  \Big)
  \nonumber\\
&+
 \int d\z' 
 \rate^{\alpha\beta}(\z|\z';t)
  \L_{\alpha}\psizp\L_{\beta}^\dagger  
-\frac{1}{2}\rate^{\alpha\beta}(\z;t)\{\L_\beta^\dagger\L_\alpha,\psiz\}_+
      \label{eq:cq-dynamics}
\end{align}
The condition that the master equation preserve probability requires
\begin{align}
\rate^{\alpha\beta}(\z;t)=\int \dz' \rate^{\alpha\beta}(\z'|\z;t)
\label{eq:J-tpcondition}
\end{align}
for each $\alpha,\beta$.
One can verify this is the case by  taking the trace of Equation \eqref{eq:J-tpcondition} and integrating over phase space, and then requiring the total probability to be conserved in time
\begin{align}
\frac{\partial}{\partial t}\int d\z\tr\psiz=0
\label{eq:tp-intime}
\end{align}
This norm preserving condition is analogous to the purely classical case of Equation \eqref{eq:rateeqnnorm} and the pure Lindbladian case.

Our derivation is general enough to allow for the coupling terms to be time-dependent, which would allow for the inclusion of various non-Markovian effects, but following our general derivation we will drop the time dependence from $\linrate^{\ag\bg}(\z;t)$ and $\rate^{\alpha\beta}(\z|\z';t)$ (one refers to such dynamics as being {\it autonomous}). 
 For the dynamics of Equation \eqref{eq:cq-dynamics} to be infinitely divisible (i.e. CP as a generator at every time $t$) it is necessary that
$\rate^{\alpha\beta}(\z|\z';t)$ be a positive semidefinite matrix in $\alpha,\beta$ $\forall \z\neq\z'$\label{par:psd}%
and  the matrix with elements $\lambda^{\ag\bg} + \rate^{\ag\bg}(\z|\z;t)$  must be PSD. These are sufficient conditions for discrete, jumping dynamics while the
conditions on $\rate^{\alpha\beta}(\z|\z';t)$ in the continuous case are derived in \cite{UCLPawula} and previewed here as Equation \eqref{eq:continuous_pos_condition}. %
For continuous quantum systems, the sum over $\alpha\beta$ is replaced by an integral, an example being the case where we take Lindblad operators to be local field operators written in terms of the coordinate $x$.
$\L_\alpha$ can be any operators, but we can always rewrite the above equation in terms of a basis of Lindblad operators orthogonal in Hilbert-Schmidt norm, such that the $\L_\ag$, (distinguished by having $\alpha$ in bold font)
are traceless, and $\L_0=\id$. %

The terms proportional to $\rate^{00}(\z|\z')$ and $\rate^{00}(\z)$ are thus purely classical and accounts for both stochastic and deterministic evolution of the classical degrees of freedom.
In the case of continuous deterministic dynamics, this term
is just the Poisson bracket $\{h^{00}(\z),\psiz\PB$
with $h^{00}(\z)$ a classical Hamiltonian.\phantomsection\label{par:pureLin}
We have in Equation \eqref{eq:cq-dynamics} separated out the  Lindbladian term with couplings $\linrate^{\ag\bg}(\z)$ to be consistent with conventions of the classical literature, but it is often convenient to include them in $\rateab$. 
The $\rate^{\alpha\beta}(\z|\z')$ can be interpreted as the jump term in both phase space or Hilbert space or both since it can induce transitions in both the classical quantum system. $\linrate^{\ag\bg}(\z)$  causes transitions in Hilbert space only, and $W^{00}(z|z')$ only transitions in phase space.
The anti-commutator terms are the 
collective no-event terms.

In the case that the quantum and classical system do not get correlated in $\z$, (for example, when one system acts as a large heat bath weakly coupled to the other) 
this reduces to a semiclassical version of the stochastic-master Equation \eqref{eq:rate-eqn} when we trace out the quantum system, while if we integrate out $\z$ we get an averaged version of GKSL dynamics. However, more generally, tracing out the quantum or classical system results in dynamics which is non Markovian, since there can be correlations between the classical and quantum systems, nor is the reduced dynamics even linear on the quantum state. When we integrate over $\z$ the presence of correlations between the classical system and the quantum one, translate into different effective coupling strengths $W(\z)$ for different quantum states \cite{alicki1995comment,vstelmachovivc2001dynamics}.\phantomsection\label{par:nonmark}

  In Section \ref{sec:Hdynamics}, we introduce a class of theories whose dynamics is close to Hamiltonian dynamics. This master equation can be expanded as
\begin{align}
    \frac{\partial\psiz}{\partial t}=
  &-i[\Hq(\z),\psiz]
 + 
  \frac{\0mom}{\tau}\Big[\L_{\alpha}\psiz\L_{\beta}^\dagger
  -\frac{1}{2}
\{\L_{\beta}^\dagger\L_{\alpha},\psiz\}_+   
\Big]
\nonumber\\
&-
\L_{\alpha}X^{\alpha\beta}_h(\z)\cdot\nabla \psiz\L_{\beta}^\dagger
+
\tau \L_{\alpha}\nabla\cdot [\friction^{\alpha\beta}_h(\z)] \psiz\L_{\beta}^\dagger
+
\tau
\L_{\alpha}
\nabla\cdot D^{\alpha\beta}(\z)\psiz\cdot\overleftarrow{\nabla}\L_{\beta}^\dagger
  +\cdots
\label{eq:weiner2}
\end{align}
where $\nabla=(\partial_\q,\partial_\p)^T$ is the divergence in phase space $\z=\q,\p$, and $X^{\alpha\beta}_h$ a matrix of vector fields which, as we will detail, is generated by a Hamiltonian 
\begin{equation}
\Hq(\z)=h\ab(\z)\L_\beta^\dagger \L_\alpha
\label{eq:decomposition}
\end{equation} over quantum and classical degrees of freedom via the symplectic matrix
\begin{align}
\Omega=\begin{bmatrix}
0 & I_n \\
-I_n & 0 \\
\end{bmatrix}
\label{eq:csymplectic}\,\, .
\end{align}
The couplings
\begin{equation} 
X^{\alpha\beta}_h(\z)=\Omega\nabla h\ab(\z)
\label{eq:HamVecField}%
\end{equation}
 are the terms which correspond to Hamiltonian evolution in the classical limit.  $\friction^{\alpha\beta}_h(\z)$ are taken to correspond to a friction term and is a vector in the $2n$ dimensional phase space at each $\z$, with components $\friction^{\alpha\beta}_{h,i}(\z)$, while the $D^{\alpha\beta}(\z)$ corresponding to diffusion terms, define a matrix both in $\alpha\beta$ and in phase space with elements
 	$D^{\alpha\beta}_{ij}(\z)$ at each $\z$.\phantomsection\label{par:components}
One typically considers $D^{\alpha\beta}(\z)$ to depend only on $\bar{q}$ and the diffusion to be in $\bar{p}$. Here, we have included the pure Lindbladian term proportional to $\linrate^{\ag\bg}(\z)$ in the $\0mom$ term.

Dropping terms of higher order in $\tau$ and taking the expectation value on both sides by performing the trace over the quantum degrees of freedom\label{pt:qtrace} gives 
a semiclassical version of the Liouville equation
\begin{align}
\frac{\partial\rho(\z;t)}{\partial t} 
=
 \tr\{\Hq(\z),\psiz\PB
+\cdots
\label{eq:semiclassical-mastereqn}
\end{align}
 If we retain the term proportional to $\tau$, an appropriate choice of $\friction\ab_h(\z)$ and $D^{\alpha\beta}(\z)$ corresponds to a hybrid version of the Fokker-Planck equation. %
 Truncating the expansion in Equation \eqref{eq:semiclassical-mastereqn} can result in negative probabilities, but the full evolution equation is completely positive and provides a natural generalisation of both classical Hamiltonian dynamics and the GKSL equation. 

We also see that taking the trace over the quantum system doesn't result in the pathological behaviour found in the semiclassical Einstein's equation \eqref{eq:semi} as depicted on the right side of Figure \ref{fig:notthis}. Instead, the dynamics of Equation \eqref{eq:semiclassical-mastereqn} is linear in the total density matrix, and similar to the situation described in our discussion of Equation \eqref{eq:homer-semi} where we have linear evolution of two coupled systems which account for the correlation. 
For example, the correlated state of two components labeled $L$ and $R$
\begin{align}
\cqstate(\z)=\frac{1}{2}\rho_L(\z)\hat\sigma_L+\frac{1}{2}\rho_R(\z)\hat\sigma_R
\end{align}
would evolve according to \eqref{eq:semiclassical-mastereqn} as%
\begin{align}
\frac{\partial \rho(\z)}{\partial t}= \frac{1}{2}\{\tr \Hq(z)\hat\sigma_L,\rho_L(\z)\}
+\frac{1}{2}\{\tr \Hq(z)\hat\sigma_R,\rho_R(\z)\}\cdots
\label{eq:semiclassical-correlated-example}
\end{align}
with the component $\rho_L(\z)$ evolving differently to  $\rho_R(\z)$ as in the left side of Figure \ref{fig:notthis}.
One can also think of $\partial/\partial p$ in Equation \eqref{eq:semiclassical-mastereqn} acting on each matrix element of $\cqstate(\z;t)$, while the velocities $\partial\Hq(\z)/\partial q$ are a matrix and thus the back-reaction force depends on which state the quantum system is in, and correlates the quantum system with the classical one.%

\section{The classical-quantum map and master equation}
\label{sec:cq-dynamics}

We will first derive the classical-quantum map which takes a cq-state, $\cqstate(\z;\t0)$ to another cq-state, state $\psizt$ and then derive the continuous time master equation. Recall that we restrict ourselves to the case of trace-class operations. We also restrict ourselves to the case where the map has an inverse at infinitesimal times (which needn't be completely positive), so that we may write the master equation in time-local form\cite{time-local}. %
Recall also Kraus's theorem, that the most general map on quantum systems $\mathcal{E}(\hat{\sigma})$ that is completely positive (CP) and trace preserving (TP) can be written so that
\begin{align}
\hat{\sigma}(t)=&\mathcal{E}(\hat{\sigma}(0))\nonumber\\
=&\sum_\mu K_\mu \hat{\sigma}(0) K^\dagger_\mu
\label{eq:kraus}
\end{align}
with the operators $K_\mu$ satisfying
\begin{align}
K_\mu^\dagger K_\mu=\id
\end{align}
where CP means that 
$\id\otimes\mathcal{E}$ acting on $\mathcal{H_S}\otimes\mathcal{H_S}$ takes positive operators to positive operators, and together with the trace preserving condition, ensures that density matrices are mapped to density matrices, even if the map acts on part of a system. 

Let us now consider the case of a map which acts on cq-states.
Because the map has to be linear in $\psiz$,%
we can write the total map
 $\X_t$ acting on $\cqstate(0)$ in terms of quantum maps $\X_{\z|\z';t}$ such that
\begin{align}
\psizt
=&\int d{\z'} \X_{\z|\z';t}\cqstate(\z';0)
\label{eq:mapsonz}
\end{align} 
\foothide{Note that Equation \eqref{eq:mapsonz} is not directly analogous to 
	the Kraus map of Equation \eqref{eq:kraus} since we write it as acting on $\cqstate(\z;t)$ as opposed to $\cqstate(t)$. In analogy with quantum theory, the former is akin to the diagonal elements of the entire matrix $\hat{\sigma}(t)$, while it is $\cqstate(t)$ (or $\cqstate_{cq}$ of Equation \eqref{eq:cq-state}) which is analogous to $\hat{\sigma}(t)$ itself. Likewise the purely classical map of Equation \eqref{eq:c-non-markovian} would be more analogous to the component form of Equation \eqref{eq:kraus}.\label{ft:totalstate} }
Since the initial state could be $\cqstate(\z';0)=\delta(\z'-\z_0)\hat{\sigma}(\z')$, i.e. a normalised quantum state with vanishing support a finite distance away from some initial point $\z_0$, the map $\X_{\z|\z';t}$ must be completely positive for all $\z,\z'$. 
The initial state which could be peaked around any $\z'$ must be mapped to another positive (but not necessarily normalised) quantum state around $\z$. \phantomsection\label{par:aroundwhat}
In the case of discrete $\z$ the maps need not preserve the trace -- they must merely be completely positive, trace non-increasing, since if we act it on the normalised state around some point $\z'$, a final state around $\z$ can't have trace larger than one as this would correspond to a probability  $\int d\z\rho(\z;t)$ larger than one. Thus one easily sees that a necessary and sufficient condition for a cq-map, is that each $\X_{\z|\z';t}$ must be a completely positive, trace non-increasing map, with the total map 
\begin{align}
\X_{\z';t}:=\int d\z \X_{\z|\z';t}
\label{eq:kraus-tp}
\end{align}
being trace-preserving. This is the analogue of the classical condition $\int d\z P(\z|\z')=1$. In the case where $\z$ are continuous, the above statements hold, except that because $\tr\psizt$ is a distribution, we should interpret statements about the non-increasing of its trace as referring to its trace smeared over an indication function $\int d\z\tr\psizt \xi(\z)$.

We can then write the map in terms of sets  of Kraus
 operators  $K_\m(\z|\z';t)$. 
 \begin{align}
  \psizt=\int d\z'\sum_{\m} K_{\m}(\z|\z';t)\varrho(\z';0)K_{\m}(\z|\z';t)^{\dagger}
  \label{eq:cq-kraus1}
\end{align}
where $\sum_\m K_{\m}(\z|\z';t)^{\dagger}K_{\m}(\z|\z';t)\leq\id$.  
This follows from the fact that the action of the entire map must preserve the trace, thus summing over $\m$ and taking the trace of both sides
of Equation \eqref{eq:cq-kraus1} on an arbitrary initial state $\varrho(\z';\t0)$, requires
\begin{align}
\sum_{\m}\int d\z\, K_{\m}(\z|\z';t)^{\dagger} K_{\m}(\z|\z';t)=\id
\label{eq:tp}
\end{align}
Equation \eqref{eq:cq-kraus1}, together with the normalisation constraint of Equation \eqref{eq:tp},
 gives the most general map on hybrid classical-quantum states. It is the hybrid generalisation of both the classical probability map of Equation \eqref{eq:c-non-markovian}, and in some sense, an extension of the Kraus representation theorem Equation \eqref{eq:kraus}. However, it can also be considered a special case of a Kraus decomposition, as we could restrict ourselves to quantum states that have the form of Equation \eqref{eq:cq-state} and write its action as
\begin{align}
\varrho_{cq}(t)=\int d\z d\z' \sum_\m\ket{\z}\bra{\z'}\otimes K_\m(\z|\z';t)\varrho_{cq}(0)\ket{\z'}\bra{\z} \otimes K_\m(\z|\z';t)^\dagger
\end{align}
where the set of Kraus operators are $\{
|\z\rangle\langle \z'| \otimes K_\m(\z|\z';t)\}$.

We now want to derive the generator $\mathcal{L}$ of this map
\begin{align}
\X_t=\exp(\int^t_0 ds \mathcal{L}_s)
\end{align}
using 
\begin{align}
\mathcal{L}_t\psizt=\lim_{\dt\rightarrow 0}\frac{\varrho(\z;t+\dt)-\psizt}{\dt}
\label{eq:genderiv}
\end{align}
This is possible, if at short times, the map has an inverse (which need not be CP), since then it can always be written in time-local form\cite{time-local} (the proof contained in \cite{buzek1998reconstruction,breuer1999stochastic} also applies to maps on CQ states). I.e. 
	\begin{align}
	\dot{\cqstate}(\z;t)=\mathcal{K}(t)\cqstate(\z';t) \label{eq:timelocal}
	\end{align}
To this end, we write the Kraus operators
in terms of an orthonormal basis of operators %
$\L_{\alpha}$ to obtain
\begin{align}
  \psizt=\sum_{\alpha,\beta=0}\int d\z'
  \X^{\alpha\beta}(\z|\z';t)\L_\alpha\varrho(\z';0)\L^\dagger_\beta
  \label{eq:cq-kraus2}
\end{align}
where complete-positivity for all $\z,\z'$ requires the $\X^{\alpha\beta}(\z|\z';t)$ be a
positive semi-definite Hermitian matrix in $\alpha,\beta$ for all $\z,\z'$ (see \cite{semigroups} and also Appendix \ref{sec:CP-conditions}). The trace-preserving condition, Equation \eqref{eq:tp} requires
\begin{align}
  \sum_{\alpha\beta}\int d\z \X^{\alpha\beta}(\z|\z';t) \L_\beta^\dagger \L_\alpha=\id
\label{eq:cq-kraus-tp}
\end{align}

Since $\{\L_\alpha\}$ is an arbitrary orthogonal basis of the operator space, we can always take $\L_0=\id$, and the orthogonality condition $\tr \L^\dagger_\beta\L_\alpha=0$ for $\alpha\neq\beta$, means that the other $\L_\alpha$ are traceless, and we denote them with bold font $\ag$. We are interested in
 changes of the state over infinitesimally short times $\dt$. 
Since the dynamics is time-local (Equation \eqref{eq:timelocal}), we
can expand the $\X^{\alpha\beta}$ as 
\begin{align}
  &\X^{00}(\z|\z';t+\delta t)= \X^{00}(\z|\z';t)-\delta(\z-\z')\pshsgo^{00}(\z';t)\dt +\rate^{00}(\z|\z';t)\dt + O(\dt^2) 
  \label{eq:x00-pre}
  \\
  &\X^{\alpha\beta}(\z|\z';t+\dt)=\X^{\alpha\beta}(\z|\z';t)+ \delta(\z-\z')
  \linrate^{\alpha\beta}(\z';t)\dt +\rate^{\alpha\beta}(\z|\z';t)\dt + O(\dt^2)\s \alpha\beta\neq 00
  \label{eq:xab-pre}
\end{align}
This is just a Taylor expansion where we have divided up the terms proportional to $\delta t$ into two contributions for later convenience, and for comparison with existing convention. %
If the map acts on an initial state $\cqstate(\z;t)$, then $\X^{00}(\z|\z';t)=\id\delta(\z-\z')$, since when $\dt\rightarrow 0$ the system has to remain at the same point in phase space and the quantum system doesn't change. The other $\X^{\alpha\beta}(\z|\z';t)=0$ since these terms  are associated with Lindblad operators $\L_\ag$ acting on the state and nothing happens to the state in the limit $\dt\rightarrow 0$. This leaves
\begin{align}
&\X^{00}(\z|\z';t+\delta t)= \delta(\z-\z')\left(1-\dt\pshsgo^{00}(\z';t)\right) +\rate^{00}(\z|\z';t)\dt + O(\dt^2) 
\label{eq:x00}
\\
&\X^{\alpha\beta}(\z|\z';t+\dt)=\delta(\z-\z')
\linrate^{\alpha\beta}(\z';t)\dt +\rate^{\alpha\beta}(\z|\z';t)\dt + O(\dt^2)\s \alpha\beta\neq 00
\label{eq:xab}
\end{align}

The first term proportional to $\dt$ in Equation \eqref{eq:xab} can be thought of as the probability per time $\delta t$ the system doesn't move in phase-space 
while the quantum system has $\L_\alpha\psiz\L_\beta^\dagger$ applied to it, while the $\rate^{\alpha\beta}(\z|\z';t)$ term in both equations gives the rate  that the system goes from $\z'$ to $\z$ and has $\L_\alpha\psiz\L_\beta^\dagger$ applied to it . We adopt the convention that the sign in front of $\gamma^{00}$ is negative anticipating that it will contribute negatively -- it's the no-event term, and the probability that the system is unchanged can only decrease with time.\phantomsection\label{par:rateconditions}

The expansion for the $\X^{00}$ component is similar to the one used to derive Equation \eqref{eq:rate-eqn} for continuous classical stochastic processes. %
Taking the limit $dt\rightarrow 0$ and keeping only the first order terms usually require that the set of operators $\L_\alpha$ be bounded in order for such terms to be small, but here we have replaced that typical assumption with the assumption that the map is invertible at short times (leading to Equation \eqref{eq:timelocal}).
To find the generator of the dynamics, we substitute Equations \eqref{eq:x00}-\eqref{eq:xab} into Equation \eqref{eq:cq-kraus2} 
to get
\begin{align}
\varrho(\z;t+\dt)
& =  \int d\z'\delta(\z-\z')\Big(\big(1-\dt\psstay^{00}(\z';t)\big)\varrho(\z';t)
  +
 \dt\linrate^{\ag 0}(\z;t)
 \L_{\ag}\varrho(\z';t)
 +
  \dt\linrate^{0\bg}(\z;t)
\varrho(\z';t) \L_{\bg}^\dagger
\nonumber\\
&+
  \dt\linrate^{\ag\bg}(\z;t)
\L_{\ag}\varrho(\z';t) \L_{\bg}^\dagger
\Big)
 + \sum_{\alpha\beta}
 \int d\z' \dt\rate^{\alpha\beta}(\z|\z';t)
  \L_{\alpha}(x)\varrho(\z';t)\L_{\beta}^\dagger
\label{eq:MEderiv1} 
\end{align}
and subtracting $\psizt$, dividing by $\dt$ and taking the limit $\dt\rightarrow 0$ gives
\begin{align}
  \frac{\partial\psiz}{\partial t}
 & =  -\psstay^{00}(\z;t)\psizt
  +
\linrate^{\ag 0}(\z;t)
 \L_{\ag}\psizt
 +
\linrate^{0\bg}(\z;t)
\psizt \L_{\bg}^\dagger
+
\linrate^{\ag\bg}(\z;t)
\L_{\ag}\psizt \L_{\bg}^\dagger
\nonumber\\
& +
 \sum_{\alpha\beta}
 \int d\z' \rate^{\alpha\beta}(\z|\z';t)
  \L_{\alpha}\varrho(\z';t)\L_{\beta}^\dagger
\label{eq:MEderiv2}  
\end{align}
As in the derivation of the GKSL equation, we now define the two Hermitian operators
\begin{align}
\Hq(\z;t):=\frac{1}{2i}(
\linrate^{0\bg}\L^\dagger_{\bg}
-\linrate^{\ag 0}\L_{\ag}
)
,\s
\NE(\z;t):=
-\psstay^{00}(\z;t)
+
\frac{1}{2}(
\linrate^{0\bg}\L^\dagger_{\bg}
+\linrate^{\ag 0}\L_{\ag}
)
\end{align}
so that we can write Equation \eqref{eq:MEderiv2} as
\begin{align}
 \frac{\partial\psiz}{\partial t}
 =
 -i[\Hq(\z;t),\psizt]
 +
 \linrate^{\ag\bg}(\z;t)\psizt
\L_{\ag}\psizt \L_{\bg}^\dagger
  + 
 \int d\z' \rate^{\alpha\beta}(\z|\z';t)
  \L_{\alpha}\varrho(\z';t)\L_{\beta}^\dagger
  +\{\NE(\z;t),\psizt\}_+
\end{align}
Having subtracted $\psizt$ from Equation \eqref{eq:cq-kraus-tp} the trace-preservation condition, Equation \eqref{eq:kraus-tp} becomes Equation \eqref{eq:tp-intime} i.e.
\begin{align}
  \int d\z\tr\frac{\partial\psizt}{\partial t}=0 
  \nonumber
  \end{align}
for any input state $\varrho(\z';t)$, which requires
\begin{align}
\NE(\z';t)=
\frac{1}{2}
\Big(
\sum_{\ag\bg}\linrate^{\ag\bg}(\z';t) \L_{\bg}^\dagger \L_{\ag}
+
\int d\z
\sum_{\alpha\beta}\rate^{\alpha\beta}(\z|\z';t) \L_{\beta}^\dagger \L_{\alpha}
\Big)
\label{eq:J-tpcondition2}
\end{align}
Defining the outgoing flux $\rate\ab(\z;t)$ by Equation \eqref{eq:J-tpcondition}, gives the desired form of the master equation, Equation \eqref{eq:cq-dynamics}.

The first term of Equation \eqref{eq:cq-dynamics}
is the free evolution of the quantum system at each point in phase-space, the second term corresponding to $\linrate\ab(\z;t)$ is pure Lindbladian, with a rate determined by the classical degrees of freedom. The remaining term, corresponding to $\rate\ab$ gives the interaction term
between the classical and quantum degrees of freedom. It is worth noting that the term corresponding to $\alpha\beta=00$
\begin{align}
\nabla^{00}\psiz:=\int d\z'\rate^{00}(\z|\z')\psizp-\rate^{00}(\z)\psiz
\label{eq:OOisLiouville}
\end{align}
is the purely classical evolution of the system and is identical to Equation \eqref{eq:rate-eqn}.
It can be stochastic, or it can correspond to deterministic dynamics generated by a classical Hamiltonian $\Hc(\z)$ in which case, we can write it in terms of the Poisson bracket, or equivalently as
\begin{align}
\nabla^{00}\psiz  =
-X^{00}_h\cdot\nabla\psiz
\label{eq:deterministicF}
  \end{align}
  where
$X^{00}_h$ is the Hamiltonian vector field 
\begin{align}
X^{00}_h=\Omega\nabla H
\label{eq:hamvec}
\end{align}
Recalling Equations \eqref{eq:HamVecField} and \eqref{eq:decomposition}
\begin{align}
X^{00}_h=(\frac{\partial h^{00}}{\partial \p},-\frac{\partial h^{00}}{\partial \q})^T
\end{align}
$\nabla^{00}$ could also include 
diffusion term as in the Fokker-Planck equation, but it will be natural to treat it in the same way we treat the other $\rate\ab$, in which case it will be a stochastic version of Hamiltonian dynamics.%

	Let us now return to Equations (\ref{eq:x00}-\ref{eq:xab}) and the condition that complete positivity places on $\X^{\alpha\beta}(\z|\z';t)$. In general, all that is required is that the total map from some initial state at $t=\t0$ to some final time is
completely positive. In this case, the generators at intermediate times $t$ need not generate completely positive dynamics if one were to replace the current state of the system $\cqstate(\z;t)$ with some other state as long as they generate CP dynamics on the state which had been evolving since $t=0$. In this way, one can consider non-Markovian dynamics. However, in the case where we demand that the generator is infinitesimally divisible, i.e. generate CP dynamics at all times on any state, then the positivity conditions for $\X^{\alpha\beta}(\z|\z';t)$ to be completely positive, can be determined from  the right-hand side of (\ref{eq:x00}-\ref{eq:xab})\phantomsection\label{par:inalphabeta}. For $\z \neq \z'$ (we can smear Equations (\ref{eq:x00}-\ref{eq:xab})  over an indicator function with support on or away from $z=z'$ as is done in \cite{UCLPawula}), $\X^{\alpha\beta}(\z|\z';t+\delta t)$ is equal to $\rate^{\alpha\beta}(\z|\z';t)\delta t$ and so it is necessary that $\rate^{\alpha\beta}(\z|\z';t)$ be a positive semi-definite matrix in $\alpha\beta$. When $\z=\z'$, we just have the usual GKSL dynamics	 and so we require $\rate^{\ag\bg}(\z|\z';t)+\linrate^{\ag\bg}(\z,t)$ be positive semi-definite\foothide{It is for this reason we use the convention of a minus sign in front of $\pshsgo^{00}(\z;t)$, but one could define $\linrate^{00}(\z;t):=1-\pshsgo^{00}(\z;t)$ in order to treat them on the same footing.}. These condition are also sufficient in the case of discrete dynamics, because these two cases completely specify the dynamics and the only condition it has to satisfy are complete positivity and norm preserving. 
The necessary and sufficient conditions on $\rate^{\alpha\beta}(\z|\z';t)$ for continuous evolution are derived in \cite{UCLPawula}.
	They are best given in terms of the moment expansion of $\rate^{\alpha\beta}(\z|\z';t)$, which is not derived until 
		Section \ref{sec:Hdynamics}, but for completeness we conclude this section with it, expressed as a matrix in $\alpha,\beta$: %
	\begin{align} \label{eq:continuous_pos_condition}
	\rate(\z|\z';t) =
	\begin{bmatrix}
	\sum_{n=0}^{2}\frac{M^{00}_n(\z',t)}{n!} \nabla_{z'}^{\otimes n}\delta(\z-\z')    & \sum_{n=0}^{1} \frac{M^{0\bg}_{n}(\z',t)}{n!} \nabla_{z'}^{\otimes n}\delta(\z-\z')  \\
	\sum_{n=0}^{1} \frac{M^{\ag 0}_{n}(\z',t)}{n!} \nabla_{z'}^{\otimes n}\delta(\z-\z')   &  M^{\ag \bg}_0(\z',t) \delta(\z-\z')  \\
	\end{bmatrix}
	\end{align}
with
	\begin{equation}
	M_2 \succeq M_1M_0^{-1} M_1^{\dag},\,\, 	
	(\mathbb{I}- M_0 M_0^{-1})M_1 =0
	\label{eq:ContinuousPosConditions}
	\end{equation} 
	Here $M^{-1}_0 $ is the generalized inverse of the PSD matrix with elements $M_0^{\ag \bg}$, $M_1$ is a matrix in both $\ag,i$ where $i$ indexes the components of $\z$. It has entries $M_{1,i}^{0 \ag}$. $M_2^{00}$ is a PSD matrix in $i,j$ with entries $M_{2,ij}^{00}$.
	This generalises the constant force dynamics of \cite{diosi1995quantum}, and gives
	Equation \eqref{eq:CQcontinuousHam}, allowing for any Hamiltonian $\Hq(\z;t)$
	to generate classical-quantum dynamics, as well as more general non-Hamiltonian back-reaction, such as those which include friction.

\phantomsection\label{par:markovian} In the remainder of the article, we will restrict our attention to the autonomous case where the coupling constants have no explicit time dependence.

\section{Non-commuting Hamiltonian dynamics}
\label{sec:Hdynamics}

We would like stochastic hybrid dynamics which  has continuous Hamiltonian evolution in the classical limit. For the continuous dynamics of Equation \eqref{eq:continuous_pos_condition}, this is given as Equation \eqref{eq:CQcontinuousHam}. In this section, we explore the discrete master equation, corresponding to Equation \eqref{eq:weiner2} whose first few terms in it's expansion 
correspond to classical-quantum versions of the Liouville equation at first order and Fokker-Planck equation at second next order.
We begin by taking $\z$ to be phase space variables $\q,\p$, and will later present the case where
$\z$ represent canonically conjugate pairs of field degrees of freedom over $x$, since the 
generalisation is straight-forward, while the notation is more cumbersome. 
The dynamics will be generated by a Hamiltonian operator which can also depend on phase space variables $\z$ and a decomposition of the 
Hamiltonian in terms of Lindblad operators $\L_\alpha$ given by Equation \eqref{eq:decomposition} which we recall as
\begin{align}
\Hq(\z)=
\hc^{\alpha\beta}(\z)\L^\dagger_\beta\L_\alpha
\nonumber
\end{align}
When we get to the field theory case, we will see that Lorentz invariance, and later, diffeomorphism invariance, place very tight constraints on this decomposition. The $h^{00}(\z)$ term is purely classical, since it couples only via $\id$ to the quantum degrees of freedom.
There is some ambiguity in what is included in the $\L_\alpha$ and what in $h^{\alpha\beta}(\z)$, but 
it will turn out that this ambiguity is inconsequential when applying this formalism to general relativity. 
The 
matrix $h^{\alpha\beta}$ is Hermitian, and it is sometimes convenient to take it to be positive as is the case for the coupling term of general relativity. 
We define
a generator of dynamics 
$X_h$ which has both phase space components and components $\alpha$,$\beta$ in Hilbert space determined by the Lindblad operators I.e.
\begin{align}
X^{\alpha\beta}_h=
\left(\frac{\partial h^{\alpha\beta}}{\partial \p},-\frac{\partial h^{\alpha\beta}}{\partial \q}\right)^T
\end{align}
Just as the Hamiltonian vector field determines the rate of flow along different directions in phase space in the case of purely deterministic classical evolution, here, $X_h^{\alpha\beta}$, additionally sets the rate of quantum jumps along different Lindblad operators.

We have already expanded $\X_t$ in terms of infinitesimally small times $\dt$, and we will now also expand it in terms of the phase space vector $\dist:=\z-\z'$, which we can then take arbitrarily small if we choose.  
To this end, let us write $\X^{\alpha\beta}(\z|\z';t)=\X^{\alpha\beta}(\z|\z-\dist;t)$ 
and perform, both an expansion in $\dt$ and a Kramers-Moyal expansion\cite{kramers1940brownian,moyal1949stochastic} in $\dist$. We can either do this directly on
Equations \eqref{eq:x00}-\eqref{eq:xab} but it's slightly less cumbersome to perform it on $\rate^{\alpha\beta}(\z|\z-\dist;t)$.
\begin{align}
\rate^{\alpha\beta}(\z|\z-\dist;t)\psizd
&=\rate^{\alpha\beta}(\z+\dist-\dist|\z-\dist;t)\psizd
\nonumber\\
&=\sum_{n=0}^\infty\frac{(-\dist\cdot\nabla)^n}{n!}\big[\rate^{\alpha\beta}(\z+\dist|\z;t)\psiz\big]
\end{align}
We now integrate this expression over $\dist$ to write
\begin{align}
\int \ddist\rate^{\alpha\beta}(\z|\z-\dist;t)\psizd
&=\sum_{n=0}^\infty(-\nabla)^{\otimes n}\cdot\big[\frac{\M{n}(\z;t)}{n!}\psiz\big]
\end{align}
in terms of the moments
\begin{align}
{\M{n}(\z;t)}=\int\ddist (\dist)^{\otimes n} \rate^{\alpha\beta}(\z+\dist|\z;t)
\label{eq:momentdef}
\end{align}
Here, we write $\M{n}(\z;t)$ as an n-fold tensor power, since $\dist^{\otimes n}\cdot\nabla^{\otimes n}=(\dist\cdot\nabla)^n$ with $\dist\cdot\nabla$ the directional divergence in the direction $\dist$. This has the advantage that the notation is compact, but the disadvantage that it distinguishes the moment corresponding to $\partial_q\partial_p$ to that of $\partial_p\partial_q$ (for example).
The zeroth moments
\begin{align}
\M{0}=\int\ddist W\ab(\z+\dist|\z;t)
\label{eq:zeroeth}
\end{align}
are just $W\ab(\z;t)$ of Equation \eqref{eq:J-tpcondition}. The first moments are the vectors
\begin{align}
\M{1}=\int\ddist \dist W\ab(\z+\dist|\z;t)
\end{align}

For a Hamiltonian $\Hq$ and decomposition $h\ab$, we will see that there is a natural choice of $\rate\ab(\z+\dist|\z;t)$ so that the vector of first moments has a contribution corresponding to Hamiltonian flow
\begin{align}
X^{\alpha\beta}_h(\z):=\Omega \nabla h\ab(\z)
\end{align}
where we have indicated that $\rate\ab(\z+\dist|\z)$ is generated from the field $h^{\alpha\beta}$, with the subscript $h$. This we do in Section \ref{ss:models}. We can also consider a contribution corresponding to friction $\friction\ab_h$ so that
\begin{align}
\int\ddist\,\dist\rate_h^{\alpha\beta}(\z+\dist|\z):=
&X^{\alpha\beta}_h(\z)- \friction\ab_h(\z)
\label{eq:firstmoment}
\end{align}
Although there could be some arbitrariness in this division, a standard friction term is of the form $\friction\ab_h(\z)=\tau\gamma\partial h\ab/\partial \p$ with $\gamma$ a constant and $\tau$ included for later convenience. 

Next there is the matrix of diffusion terms
\begin{align}
D^{\alpha\beta}(\z):=\frac{1}{2}\int \ddist\dist\otimes\dist\rate_h^{\alpha\beta}(\z+\dist|\z).
\label{eq:secondmoment}
\end{align}

Making the Markovian approximation so that we may drop the explicit dependence on time, and putting this together gives
\begin{align}
\int \ddist\rate^{\alpha\beta}(\z|\z-\dist)\psizd
&=\rate_h(\z)\psiz-X_h\ab\cdot\nabla\psiz
+
\nabla\cdot\friction_h\ab\psiz
+
\nabla\cdot D^{\alpha\beta}(\z)\psiz\cdot\overleftarrow{\nabla}
+ \cdots
\label{eq:W-KM}
\end{align}
where we have used the fact that $\nabla\cdot X_h^{\alpha\beta}=0$,
and the notation $\overleftarrow{\nabla}$ is to indicate that $\nabla$ acts on the second $\dist$ in the tensor product of $D^{\alpha\beta}(\z)$, while $\nabla$ acts on the first. One typically refers to the terms which contain $\nabla\psiz$ as the {\it drift}, since they determine the rate at which the expectation values of observables change. The terms corresponding to second derivatives lead to diffusion in phase space. Note that if $D^{\alpha\beta}(\z)$ depends on both the $q$ and $p$, then it can also include contributions to the drift, sometimes refered to as {\it anomolous drift}. Here we will take $D^{\alpha\beta}(\z)$ to only depend on the $q$ so that there are no anomalous drift terms coming from this and higher order terms.

The Pawula theorem\cite{pawula1967rf}, says that either the Kramers-Moyal expansion truncates at the second moment, or requires an infinite sum of terms.  In the latter case, our truncated master equation will not be completely positive, and should be regarded as an approximation of the dynamics of Equation \eqref{eq:cq-dynamics}.
We will call a particular expansion of $\Hq$ in terms of Lindblad operators, and the choice of $\rate^{\alpha\beta}(\z+\dist|\z)$ or $\X^{\alpha\beta}(\z+\dist|\z)$ a {\it realisation} 
and although we will present two simple realisations in Section \ref{ss:models},
the main requirement we demand is that the moment expansion include $X^{\alpha\beta}_h(\z)$, the Hamiltonian vector field for $h^{\alpha\beta}(\z)$ in it's first moment, Equation \eqref{eq:firstmoment}. This ensures that we are able to reproduce the deterministic classical equations of motion in the appropriate limit. When we consider applying our results to field theory and later general relativity,  Lorentz covariance and parameter invariance implies further constraints on $\rate\ab(\z|\z')$. 

Substituting the Kramers-Moyal expansion of $\rate^{\alpha\beta}(\z|\z')$ into Equation \eqref{eq:cq-dynamics}
gives 
\begin{align}
  \frac{\partial\psiz}{\partial t}=
  &-i[\Hq(\z),\psiz]
 + 
 \big(\linrate\ab(\z) +
  \M{0}(\z)\big)\Big[\L_{\alpha}\psiz\L_{\beta}^\dagger
  -\frac{1}{2}
\{\L_{\beta}^\dagger\L_{\alpha},\psiz\}_+   
\Big]
\nonumber\\
&-
\L_{\alpha}X^{\alpha\beta}_h(\z)\cdot\nabla \psiz\L_{\beta}^\dagger
+ 
\L_{\alpha}\nabla \cdot[\friction\ab_h(\z)\psiz]\L_{\beta}^\dagger
+
\L_{\alpha}
\nabla\cdot D^{\alpha\beta}(\z)\psiz\cdot\overleftarrow{\nabla}\L_{\beta}^\dagger
  +\cdots
  \nonumber
\end{align}
When we consider realisations of $\rate\ab$, we typically find that each order in the expansion carries with it higher powers of some constant $\tau$. If we identify
$\frac{1}{\tau}\rate\ab(\z):=\linrate\ab(\z)+\M{0}$, and $D\ab\rightarrow \tau D\ab$ this is Equation \eqref{eq:weiner2} highlighted in the Introduction.

An important category of realisations, are ones in which $\M{n}(\z)$ for $n\geq 2$ and $\friction\ab_h(\z)$ only depend on the conjugate coordinates $\q$, and the diffusion is only in the conjugate momenta $\p$. I.e.
\begin{align}
\frac{\partial\psiz}{\partial t}=
&-i[\Hq(\z),\psiz]
+ 
\frac{\0mom}{\tau}\Big[\L_{\alpha}\psiz\L_{\beta}^\dagger
-\frac{1}{2}
\{\L_{\beta}^\dagger\L_{\alpha},\psiz\}_+   
\Big]
\nonumber\\
&-
\L_{\alpha}X^{\alpha\beta}_h(\z)\cdot\nabla \psiz\L_{\beta}^\dagger
+
\tau \L_{\alpha}\frac{\partial}{\partial p}\cdot[ \friction^{\alpha\beta}_h(\z) \psiz]\L_{\beta}^\dagger
+
\tau
\L_{\alpha}\frac{\partial}{\partial p}
\cdot D^{\alpha\beta}(\z)\psiz\cdot\overleftarrow{\frac{\partial}{\partial p}}\L_{\beta}^\dagger
+\cdots
\label{eq:weiner2-onlyp}
\end{align}
This means that the $\q$ still obey the equation of motion
\begin{align}
\dot{\q}=\frac{\partial h^{00}}{\partial \p}
\end{align}
and allows one to define $\p$ deterministically in terms of $\q$ and $\dot{\q}$. This can be verified using the Heisenberg representation for all moments of $\q$ (see Section \ref{sec:heisenberg}).
Equation \eqref{eq:weiner2-onlyp} is thus a classical-quantum version of the Wiener process or Fokker-Planck equation and we will refer to such a master equation as {\it Brownian}. For such a master equation, it is the conjugate momentum which is influenced by the quantum system. In the more general
case, $\p$ is only defined in terms of expectation values. 

It's worth noting that in our expansion, we have a purely Lindbladian term which is itself trace-preserving, and a total divergence term, with the quantity $X^{\alpha\beta}_h-\friction\ab_h- D^{\alpha\beta}\cdot\overleftarrow{\nabla}$ acting as a current. Thus if we integrate over phase space, this term also conserves probability, contributing only a boundary term. Generally though, this term is not positive by itself, and one typically has to include the full expansion to ensure this.
Looking at each of the terms in Equation \eqref{eq:weiner2}, we see that the first term is just the deterministic quantum evolution. One can have a purely deterministic classical evolution term as well. Since $\L_0=\id$, the $X_h^{00}$ term is generated by the pure classical term $h^{00}$ in the expansion of $\Hq(\z)$ in Equation \eqref{eq:decomposition}.
\begin{align}
-X^{00}_h(\z)\cdot\nabla\psi=\{h^{00}(\z),\psiz\PB
\end{align}
The other deterministic term, at least on the classical system and for a particular $\alpha\beta$ is $\L_\alpha X_h\ab(\z)\cdot\nabla\psiz\L^\dagger_\beta$, which, if we take the trace over the quantum system, gives the required dynamics we are looking for, Equation \eqref{eq:semiclassical-mastereqn}. 

The other non-deterministic parts include the pure Lindblad term, and if we trace out the quantum system it contributes nothing to the dynamics of the classical system, except indirectly, as it evolves the quantum system which back-reacts on the classical system.  On the quantum system it can generate decoherence, and if the Lindblad operators are diagonal in raising and lowering operators, it raises and lowers the state of the quantum system. The final two terms, if we trace out the quantum degrees of freedom are the friction and diffusion terms on phase space, reminiscent of the Fokker-Planck equation, although here, there is again some non-Markovianity as the quantum system can serve as a memory for the classical system.

The Lindbladian term can determine the rate at which the wave-function decoheres and if this rate is small then the diffusion term tends to be large. This is to be expected since as discussed in the introduction, a purely deterministic evolution on phase space would decohere the wave function instantly, and so a slow rate of decoherence requires either a very noisy stochastic evolution or a very slow jump rate.

It may be worth noting that we can now write a formal solution of Equation \eqref{eq:weiner2}. One way to do this, is to  double the Hilbert space\cite{FeynmanVernon1963,jamiolkowski1972linear,choi1975completely}\foothide{For the moment we take the quantum system to be finite dimensional, since extension to the continuous case is straightforward} to write the density matrix $\psiz$ in its eigenbasis
\begin{align}
\psiz=\sum_i p_i(\z)\proj{i}
\end{align}
as a pure entangled state in the Schmidt basis
\begin{align}
\ket{\psiz}_{AB}=&\sum_i \sqrt{p_i(\z)}\ket{i}_A\otimes\ket{i}_B
\end{align}
The CQ-master Equation of \eqref{eq:weiner2-onlyp} can then be written as
\begin{align}
\frac{\partial\ket{\psiz}_{AB}}{\partial t}
=&\Big[-i\Hq(\z)\otimes\id+i\id\otimes\Hq^T(\z)
+(\linrate\ab(\z)+\rate\ab(\z)) \big(\L_{\alpha}\otimes\L^*_{\beta}
-\frac{1}{2}\L^\dagger_\beta\L_\alpha\otimes\id
-\frac{1}{2}\id\otimes\L^T_\alpha\L^*_\beta\big)
\nonumber\\
&-
\L_{\alpha}\otimes\L^*_{\beta}X_h\ab\cdot\nabla
+
\L_{\alpha}\otimes\L^*_{\beta}\nabla \cdot\friction\ab_h(\z)
+
\L_{\alpha}\otimes\L^*_{\beta}\,\nabla^{\otimes 2}\cdot D^{\alpha\beta}(\z)
+\cdots
\Big]\ket{\psiz}_{AB}
\label{eq:doubled}
\end{align}
where we have used the fact that applying an operator to one subsystem of the entangled state, is equivalent to applying the transpose to the other subsystem\cite{jozsa1994fidelity}.
If we designate the operator in square brackets as $\bm{\mathcal{L}}$ then the formal solution to the rate equation is
\begin{align}
\ket{\psizt}_{AB}=e^{\bm{\mathcal{L}}t}\ket{\cqstate(\z;0)}
\label{eq:prop}
\end{align}
The generator on the doubled system is in some sense a Hamiltonian, albeit one with an imaginary part.

\subsection{Realisations of $\rate^{\alpha\beta}(\z,\z-\dist)$}
\label{ss:models}

Here, we present some simple and constructive examples for $\rate^{\alpha\beta}(\z,\z-\dist)$ which have the properties we require, namely, positive semi-definite matrices whose first moment contain a term which generates Hamiltonian flow via Equation \eqref{eq:firstmoment}.
A simple realisation, valid for small $\tau$ is
\begin{align}
\rate^{\alpha\beta}(\z,\z-\dist)
&\approx
\frac{1}{\dtau}
\delta^{(2N)}(\dist-\dtau X^{\alpha\beta}_h(\z))
\label{eq:simplemodel}
\end{align}
This gives the Liouville term as the first moment as required, and $1/\tau$ as the zeroth moment governing the collapse rate. We could take 
$\dtau$ to depend on $\alpha,\beta$, or include a distribution $f^{\alpha\beta}$ over $\tau$. %
The realisation of Equation \eqref{eq:simplemodel} can be considered an approximation of
\begin{align}
\int\ddist\rate^{\alpha\beta}(\z,\z-\dist)\psizd
=
\frac{1}{\dtau}
e^{%
	\dtau\{h\ab,\cdot\}}\psiz
\label{eq:alsosimplemodel}
\end{align}
with the right-hand side being the exponential of the Poisson bracket evolution, $\{h\ab(\q,\p),\cdot\}:=\sum_k\Big[\frac{\partial h^{\alpha\beta}}{\partial q_k}\frac{\partial}{\partial p_k}
-
\frac{\partial h^{\alpha\beta}}{\partial p_k}\frac{\partial}{\partial q_k}\Big]$
which generates a finite shift in phase space along the flow generated by $h$ (dropping the $\alpha\beta$ index for simplicity). This is the classical analogue of the evolution operator induced by the quantum Hamiltonian $\hat{H}$, $e^{-i\hat{H} t}$ since\phantomsection\label{par:expPoisson}
	\begin{align} 
e^{\tau\{h,\cdot\}}
\cqstate(q,p;\t0)
	&=	\varrho(q-\int_\t0^\tau dt\frac{\partial h}{\partial p},p+\int_\t0^\tau dt\frac{\partial h}{\partial q};\t0)\nonumber\\
	&\approx \varrho(q-\dtau\frac{\partial h}{\partial p},p+\dtau\frac{\partial h}{\partial q};\t0)
	\label{eq:finiteshift}
	\end{align}
Then $e^{\dtau\{h,\cdot\} }-1$ acts like a finite time divergence along the direction of Hamiltonian flow generated by $h$.

This lead to a natural non-commuting generalisation of this directional divergence which for a CQ Hamiltonian with a single Lindblad operator $\Hq=h\L^\dagger\L$, and for small $\tau$, is
\begin{align}
\boldsymbol{\nabla}\cqstate(q,p):=
\frac{1}{\tau} 
e^{\tau\{h,\cdot\}}
\L^\dagger
\cqstate
\L-\frac{1}{2\tau}\{\L^\dagger\L,\cqstate\}_+
\label{eq:simplenoncomdivergence}
\end{align}

In
the case of many Lindblad operators let us define basis vectors $\baseab$ and define
\begin{align}
\PQdiv\psiz\baseab
:=\frac{1}{\tau}\Big[
\int\ddist\rate_h\ab(\z|\z-\dist)\L_\alpha\psizd\L_\beta^\dagger
-\rate_h\ab(\z)\frac{1}{2}\{\L^\dagger_\beta\L_\alpha,\psiz\}_+
\Big]
\label{eq:PQdiv}
\end{align}
leading to
\begin{align}
\PQdiv\psiz\baseab =
\frac{
\L_\alpha e^{
	\tau
\{h\ab,\cdot\PB
}
\psiz \L^\dagger_\beta
-\frac{1}{2}\{\L^\dagger_\beta\L_\alpha,\psiz\}_+
}{\dtau}
\label{eq:PQdiv-simple}
\end{align}
which for small $\tau$ looks very much like  a finite directional divergence (both along the direction $X_h\ab$ and along the Lindblad operators). 

This definition depends on the decomposition of $\Hq$ into Lindblad operators, but one can first diagonalise $\hab$, in which case we will write the left-hand side of Equation \eqref{eq:PQdiv-simple} as ${\boldsymbol{\nabla}}_{\Hq}\cqstate(\z;t)$.
In cases where
\begin{align}
\PQdiv\psiz\baseab\approx\frac{1}{\tau}\Big[
\L_\alpha\cqstate(\z-\tau X\ab_h)\L^\dagger_\beta
-
\frac{1}{2}\{ \L^\dagger_\beta\L_\alpha,\psiz  \}_+
\Big]
\label{eq:non-commuting_divergence}
\end{align}
it is sometimes the case that $X\ab_h$ is non-negative (or non-positive), in which case this discrete, non-commuting directional derivative takes the simpler form of a non-commuting finite difference
\begin{align}
X\ab_h
\PQdiv\psiz\approx\frac{X\ab_h}{\tau}\Big[
\L_\alpha\cqstate(\z-\tau)\L^\dagger_\beta
-
\frac{1}{2}\{ \L^\dagger_\beta\L_\alpha,\psiz  \}_+
\Big]
\label{eq:non-commuting_divergence-lin}
\end{align} Care must be taken to ensure it is norm preserving, which is the case if for example $X\ab_h$ depends only on $q$ and the jump is in $p$. Although
Equation \eqref{eq:PQdiv-simple} looks like a discrete CQ Poisson bracket,
${\boldsymbol{\nabla}}_{\Hq}\cqstate(\z;t)$ is only linear in $\cqstate(\z;t)$ and not $\Hq$. On the other hand, the right-hand side of Equation \eqref{eq:alex}, which
uses the Aleksandrov-Gerasimenko bracket is bilinear and so
does have some properties more akin to a novel bracket although its action is not CP. The same holds true if we replace the Aleksandrov-Gerasimenko bracket in Equation \eqref{eq:alex} with the 0th and first moment of %
${\boldsymbol{\nabla}}_{\Hq}\cqstate(\z;t)$. In some cases, we can have a bilinear map which is also CP as in Equation \eqref{eq:non-commuting_divergence-lin}.\label{ft:cqbracket}

The non-commuting divergence is trace-preserving and completely positive, and allows us to write the master equation as
\begin{align}
\frac{\partial\psiz}{\partial t}=
  &-\frac{i}{\hbar}[\Hq(\z),\psiz]
+\PQdiv\psiz\baseab
  \nonumber\\
  =& -\frac{i}{\hbar}[\Hq(\z),\psiz]
  +\frac{1}{\tau}
  \Big[
\L_\alpha\psiz\L^\dagger_\beta
-
\frac{1}{2}\{ \L^\dagger_\beta\L_\alpha,\psiz  \}_+
\Big]
 -X_h\ab\cdot\L_\alpha \nabla\psiz\L_\beta^\dagger
  +\cdots
\label{eq:simplemaster}
\end{align}
If we make the model more deterministic ($\tau\rightarrow 0$), the strength of the pure-Lindbladian term increases to compensate, while if we make the Lindbladian term small, the diffusion terms at second order become significant. Here, we have temporarily switched units to show $\hbar$, since it strongly suggests that we set $\tau=\hbar$. Thus as $\hbar\rightarrow 0$, the decoherence renders the system completely classical, the diffusion terms and those of higher order go to zero, and we are left with purely continuous classical equations of motion. For gravity, the realisations are more 
tightly constrained due to diffeomorphism invariance, but similar realisations can be used to obtain dynamics which recover general relativity in the classical limit. 

\todo{I have edited until here}

\subsection{Example: Qubit coupled to a classical potential}
\label{sec:qubit}

Let us now review a simple system, solved in \cite{UCLqubit} with Šoda, Sparaciari and Weller-Davies, which will contain some features we will encounter with gravity\cite{carlo_foot}.  While the model bears some similarity to the Stern-Gerlach experiment (c.f. also \cite{diosi2000quantum}), the behaviour of the spin can differ depending on how one decomposes the Hamiltonian in terms of Lindblad operators. 
Our system will consist of a classical particle which will couple to the spin of a two level system or the energy of a quantum harmonic oscillator. The classical degrees of freedom are the position $q=x$ and momentum $p$ of the particle, with free evolution given by $h^{00}=p^2/2$. We could take the purely classical evolution \phantomsection\label{par:pureclassical}coming from $\rate^{00}(\z|\z';t)$ to be deterministic or stochastic and we will consider the former here and discuss the latter in \cite{UCLqubit}.  We take the total cq-Hamiltonian to
be given by $\Hq(\z)=p^2/2m+B(q)\Hqm$ with $\Hqm=\omega \frac{1}{2}({\bfa}^\dagger{\bfa}+{\bfa}{\bfa}^\dagger)$ for the oscillator and 
$\Hqm=
\begin{pmatrix} \omega & 0\\ 0 & -\omega\end{pmatrix}$ for the qubit.

To model a Stern-Gerlach experiment we consider a two level system with states $\ket{\up}$,$\ket{\down}$ and a cq-density matrix which depends on the classical degrees of freedom.
\begin{align}
\varrho(q,p)=
\begin{pmatrix} u_\up(q,p) & \alpha(q,p)\\ \alpha^*(q,p) & u_\down(q,p)\end{pmatrix}
\label{eq:cqQubit}
\end{align}
For simplicity we consider a linear potential $B(q)=Bq$ and a decomposition of the Hamiltonian, Equation \eqref{eq:decomposition}, in terms of Lindblad operators 
 $\L_\up=\sqrt{\omega}\ket{\up}\bra{\up}$, $\L_{\down}=\sqrt{\omega}\ket{\down}\bra{\down}$ with $h^{\up\up}(\z)=-h^{\down\down}(\z)=Bq$ and the others zero.
An example which builds toward understanding the field theoretic case is the harmonic oscillator with Lindblad operators $\bfa$ and $\bfa^\dagger$. This is more interesting in the sense that the pure Lindbladian term raises and lowers the state of the quantum system. In the qubit system this corresponds to choosing Lindblad operators
$\L_\up=\sqrt{\omega}\ket{\down}\bra{\up}$, $\L_{\down}=\sqrt{\omega}\ket{\up}\bra{\down}$ with diagonal couplings $\rate^{\ag\bg}(\z|\z-\dist)=\delta_{\ag\bg}\rate^{\ag\bg}(\z|\z-\dist)$, or even a single Lindblad operator
$\L=\begin{pmatrix} 0 & \sqrt{\omega}\\ \sqrt{\omega} & 0\end{pmatrix}$.
Here, we just present the more trivial first case decomposition -- the other cases, including non-trivial $B(q)$ are studied both analytically and numerically using an unravelling approach in \cite{UCLqubit}.
We take a simple realisation of the discrete dynamics, that of Equation \eqref{eq:simplemodel} for small $\tau$.
\begin{align}
\frac{\partial\varrho(q,p)}{\partial t}
 =& -i[\Hqm(q,p),\varrho(q,p)]+\{h^{00},\varrho(q,p)\PB
+\sum_{\alpha,\beta}
\frac{1}{\tau}\Big[
\L_\alpha 
\varrho(q-\tau\frac{\partial h^{\alpha\beta}}{\partial p},p+\tau\frac{\partial h^{\alpha\beta}}{\partial q})
 \L^\dagger_\beta
-\frac{1}{2}\{\L^\dagger_\beta\L_\alpha,\varrho(q,p)\}_+
\Big]
\nonumber
\\
=&
-iBq[\begin{pmatrix} \omega & 0\\ 0 & -\omega\end{pmatrix},\varrho(q,p)]
+\{\frac{p^2}{2m},\varrho(q,p)\PB
+
\frac{\omega}{\tau}
\proj{\up}\varrho(
q,p+\tau B)
 \proj{\up}
\nonumber\\ 
& 
+\frac{\omega}{\tau}\proj{\down}
\varrho(
x,p-\tau B)
 \proj{\down}
-\frac{\omega}{2\tau}\{
\id,\varrho(q,p)\}_+
\label{eq:qubitinpot}
\end{align}

In terms of the matrix elements of the cq-state, Equation \eqref{eq:qubitinpot}, the master equation decouples into two equations for the diagonal components and one for  the off-diagonal coherence term
\begin{align}
\frac{\partial\u_\up(q,p)}{\partial t} & =-\frac{p}{m} 
 \frac{\partial\u_\up(q,p)}{\partial q}
 +\frac{\omega}{\tau}\big[
 \u_\up(q,p+\tau B)-\u_\up(q,p)\big]
\nonumber\\
 \frac{\partial\u_\down(q,p)}{\partial t} & =-\frac{p}{m} 
 \frac{\partial\u_\down(q,p)}{\partial q}
 +\frac{\omega}{\tau}\big[
 \u_\down(q,p-\tau B)-\u_\down(q,p)\big] 
  \label{eq:qubit-diag}\\
 \frac{\partial\alpha(q,p)}{\partial t} &=
-2iBq\omega\alpha(q,p) 
 -\frac{p}{m} 
 \frac{\partial\alpha(q,p)}{\partial q}
 -\frac{\omega}{\tau}\alpha(q,p)
 \label{eq:qubit-offdiags}
\end{align}
The latter has a solution
\begin{align}
\alpha(q,p)=\rho(q-\frac{p}{m}t)e^{-2iB\omega t(q-\frac{pt}{2m}) -\frac{\omega t}{\tau}}
\end{align}
where $\rho(q-pt)$ is any normalised function of its argument. We see that the coupling of the qubit results in decoherence, since the off-diagonal elements decay exponentially fast. 

More than that, the wave function collapses into a definite state of being in the up or down state.
This we see through the equation for the diagonal terms. Comparing Equations \eqref{eq:qubit-diag} to Equation \eqref{eq:rate-eqn}, we see that at a rate of $\omega/\tau$ the system undergoes jumps in momentum by $\pm\tau B$ depending on its state.  Since we can monitor the classical degrees of freedom without disturbing the system, measuring the change in momentum will uniquely determine the value of the spin. In this sense, the interaction leads to more than just decoherence, but an objective collapse of the wave-function\cite{shimony1990desiderata,sep-qm-collapse}. 
While the basis to which the quantum state collapses can appear arbitrary in spontaneous collapse models, here it is determined by the form of the CQ-interaction which will be tightly constrained by the requirement of reproducing Einsten gravity in the classical limit. More generally, if the backreaction force $\nabla h\ab$ 
is unique when  diagonalised, then monitoring the classical system uniquely determines which Lindblad operators were applied to the state. Thus an initially pure quantum state, remains pure conditioned on the classical degrees of freedom. In the continuous realisations of the theory, this has since been shown to be generically true in \cite{layton2022semi} with Layton and Weller-Davies, provided an inequality we call the {\it decoherence-diffusion trade-off} is saturated.\label{ft:unique}

The solution (a Poisson distribution) for the diagonal terms of $\varrho$ in this and a more general setting is given in \cite{UCLqubit} but the evolution can be understood in broad terms by  inspection. The particle is making a momentum jump of $\pm\tau B$ at a rate of $\omega/\tau$, so on average it has an acceleration  of $\pm\omega B$ exactly as we would expect for a Stern-Gerlach device. We can imagine that at each $dt$, there is a probability of the particle undergoing a jump in momentum, so by the law of large numbers, we expect the particle to have made on average $\omega t/\tau$ jumps, normally distributed with a variance of $\frac{\omega}{\tau}(1-\frac{\omega}{\tau})$.
If we Taylor expand Equation \eqref{eq:qubit-diag}
\begin{align}
\frac{\partial\u(q,p)}{\partial t} & =-\frac{p}{m} 
 \frac{\partial\u(q,p)}{\partial x}
 \pm\omega B\frac{\partial\u(q,p)}{\partial p}
 +\frac{\omega B^2 \tau}{2}\frac{\partial^2\u(q,p)}{\partial p^2}
+\cdots
\end{align}
(where $u$ is $u_\up$ or $u_\down$, for $\pm$ respectively), then to first order we have a particle undergoing acceleration $\pm \omega B$ depending on its spin, as expected, with a diffusion term at higher order. However, care must be taken, as if we truncate the Taylor expansion, the density matrix can become negative. Equation \eqref{eq:qubitinpot} is completely positive however, and can be understood as follows: an initial pure state in superposition will initially have a constant momentum, but will then undergo jumps in momentum of finite size $\tau B$, and at a rate 
$\omega /\tau$. This causes the state to collapse to the up or down state, at a rate $\omega/\tau$. 

If we trace out the qubit and look at the average value of the particle's trajectory, then for a uniform superposition of up and down spin, it will have zero average acceleration, but if we look at the full state, the particle undergoes acceleration (albeit stochastically) in a direction which depends on the value of the spin. 
We thus see that the evolution of the particle corresponds to the left panel of Figure \ref{fig:notthis}, not the right one.

\subsection{Post-quantum field theory}
\label{ss:PQFT}

In order to consider applying our master equation to field theories, we must first ensure that it can be made Lorentz covariant.
Although it was initially thought that such dynamics are impossible\cite{srednicki-purity}, GKSL equations which are covariant under the proper orthochronous Lorentz group were introduced in \cite{alicki-reldecoherence}. This was further taken up in \cite{poulinKITP,OR-intrinsic} (see also \cite{beckman2001causal}).  A Lorentz invariant Schwinger-Keldish action for open quantum scalar field theory was used to show that the GKSL equation for scalar field theories is renormalizable\cite{baidya2017renormalization}. The formalism used there is manifestly Lorentz invariant. There are however subtleties that have since been clarified with Weller-Davies in \cite{UCLLorentz}. The full master equation of \cite{baidya2017renormalization} is not completely positive, but one can find Lorentz invariant master equations which are CP, Lorentz invariant and renormalisable, as well as master equations which are CP, renormalisable and Lorentz covariant, but not Lorentz invariant. The Lindbladian corresponding to the discrete realisation presented in Section \ref{sec:PQgrav} is covariant but not invariant, while the Lindbladian of the continuous realisation is invariant but not covariant. This apparent paradox is resolved by showing that transformations of space-time are not generated by a unitary, but rather, by the semi-group inherited from the Lindbladian, and thus there are different notions of invariance and covariance\cite{UCLLorentz}. Here, we focus on covariance conditions, and address the subtleties in more detail in \cite{UCLLorentz,oppenheim2023covariant}\label{ft:renorm}.

In order to make the dynamics here Lorentz-covariant, one can begin by choosing Lindblad operators and couplings, such that the right-hand side of Equation \eqref{eq:cq-dynamics} transforms like $\partial/\partial t$, so that both sides of the equation transform
in the same way under a Lorentz boost.  This has been referred to as a {\it minimal Lorentz covariance} requirement\cite{srednicki-purity}.
We will first construct a field theory version of Equation \eqref{eq:cq-dynamics} and then discuss how to make it covariant under the proper orthochronous Lorentz group.

To obtain the field theoretic version of the master equation, we take the Lindblad operators $\L_{\alpha}$ to also depend on the spatial coordinate $x$ and in particular, take them to be local field operators  $\L_{\alpha}(x)$. We can think of them as carrying a double index $\alpha,x$. This requires the couplings $\rate$ to carry not just an index $\alpha\beta$ but also $x,y$.
Then, instead of a sum or integral over just $\alpha,\beta$, we also have an integral over $x$,$y$, and write the couplings as $\rateabx$ with the phase space degrees of freedom $\z$ being local fields at each point $x$.

Cluster-separability\cite{cluster_foot} demands that 
the coupling constants $\rateabx$ be delta functions in $x,y$\cite{bps}, and that the coupling to phase space variables $\z$ be to the local classical field $\z(x)$. A violation of cluster-separability can result in violations of Lorentz invariance, with correlations or entanglement being created over space-like separations. However, this need not violate causality or allow for superluminal signalling\cite{OR-intrinsic}. Thus, while it is desirable to have $\rateabx$ be delta functions,  for greater generality we will allow the coupling constants $\rateabx$ to have support at finite $|x-y|$. Should these couplings not fall off fast enough with increasing $|x-y|$, the theory is likely to run afoul of experiment.
For the cq-state at $\z$, we will continue to write it as $\psizt$ with the understanding that it lives in Fock space, and is an operator-valued distribution functional of the classical fields $\z(x)$ over the entire manifold of points $x$. If we consider the points $x$ to be on a lattice, with the continuum obtained in the limit that the lattice spacing goes to zero, then we can think of the measure ${\cal D}\z$ over phase space as $\Pi_xdq(x)dp(x)$, and in the continuum limit, the master equation is given by a functional integral.

If we take the Lindblad operators to be local field operators $\lax$, then the field theoretic version of Equations \eqref{eq:cq-dynamics} and \eqref{eq:weiner2} is 
\begin{align}
  \frac{\partial\psiz}{\partial t}
  &=-i[\Hq(\z),\psiz]
 + \int dxd^3y\ddf
\rateabxd
\L_{\alpha}(x)\psizd\L_{\beta}^\dagger(y)
  -
\frac{1}{2}\rate_h\ab(\z;x,y)\{\lby\lax,\psiz\}_+   
\\
  &=-i[\Hq(\z),\psiz]
 + 
 \int dx d^3y
\rate_h\ab(\z;x,y)\Big[ \lax\psiz\lby
  -\frac{1}{2}
\{\L_{\beta}^\dagger(y)\L_{\alpha}(x),\psiz\}_+   
\Big]
\nonumber\\
&
-
\int dx  
\Big[  X^{\alpha\beta}_h(\z;x)
\lax\cdot\nabla \psiz\lbx
- \nabla\cdot D_h^{\alpha\beta}(\z;x,y)\cdot
\L_{\alpha}(x)
\psiz
\L_{\beta}^\dagger(y)
\overleftarrow{\nabla}\nonumber\\
&
-\int dx\nabla\cdot[
\friction^{\alpha\beta}_h(\z;x,y)
\lax\psiz\lby]
  \Big] +\cdots  
\label{eq:weinerfields}
\end{align}
One can also consider master equations which contain multiple jumps at higher order, i.e.
	\begin{align}
		\int \ddf dx_1 dx_2... \rate^{\alpha_1\beta_1,\alpha_2\beta_2...}(\z|\z-\dist;x_1,x_2,...)\cdots\L_{\alpha_2}(x_2)\L_{\alpha_1}(x_1)\psizd\L^\dagger_{\beta_1}(x_1)\L^\dagger_{\beta_2}(x_2)\cdots
		\nonumber
	\end{align}  but we will not do so here.

In flat space, one can consider as an example the scalar field discussed later in Section \ref{sec:PQgrav}, and take the pure Lindbladian coupling term $\rate\ab(\z;x,y)$ to be a close to a delta function, and proportional to $1/\tau$.
If we wish to reproduce Hamiltonian dynamics in the classical limit, then we will generally take the $X_h\ab$ term to transform just as the Poisson bracket would for a Lorentz-invariant classical field theory, since the only difference between it and $\{\Hq(\z),\psiz\PB$ is operator ordering.

 As an example 
we could decompose the Hamiltonian of the field theory in terms of the Lindblad operators used in \cite{alicki-reldecoherence,poulinKITP,baidya2017renormalization}
\begin{align}
\Hq(x)=\int dx\habx\lbx\lax
\label{eq:fieldham}
\end{align} 
with $\habx$ coupling the quantum field to a classical field $\z(x)$, which in the case of gravity, we can take to be the 3-metric and its conjugate 3-momentum. The couplings
\begin{align}
X_h\ab(\z;x)=\frac{\delta h\ab(\z;x)}{\delta \z(x)}
\end{align}
then reproduce the Hamiltonian back-reaction of the quantum fields on the classical ones.
If one wants to be explicit about $\rateabx$ one can use the realisation presented in Section \ref{sec:PQgrav}. 

In \cite{UCLPawula} we show that depending on the decomposition $\habx$, there are discrete dynamics or continuous dynamics. In particular, it is shown that the Kramers-Moyal expansion in Equation \eqref{eq:weinerfields} never terminates and leads to finite sized jumps in phase space unless the only non-zero $D_h^{\alpha\beta}(\z;x,y)$ are $D_h^{00}(x,y)$, and the only non-zero $X_h^{\alpha\beta}(\z;x)$ are $X_h^{\alpha 0}(\z;x)$ and  $X_h^{0\alpha}(\z;x)$, and the higher order terms all vanish. In other words, the continuous master equation is obtained by restricting the decomposition of Equation \eqref{eq:fieldham} to only include terms
\begin{align}
\Hq(x)=\int dx h^{0\bf{\alpha}}(\z;x)\id\L_{\bf{\alpha}}(x)
\end{align}
while otherwise, one has discrete dynamics.

The positivity conditions in the field theoretic case are analogous to the finite dimensional conditions of Equation \eqref{eq:ContinuousPosConditions}. Recalling that for $2n$ phase space degrees of freedom per point, $D_{h}^{00}(\z;x,y)$ is a $2n\times 2n$ matrix over $x,y$ and denoting $\rate_h(\z;x,y)$ as the matrix with elements  $\rate_h\ab(\z;x,y)$, and $X_h(\z;x)$ as the vector with elements $X_h^{0\alpha}(\z;x)$, the continuous master equation is completely positive
provided that the square matrix
\begin{align}
\label{eq:positivity_continuous_field}
\begin{bmatrix}
D_h^{00}(x,y) & \delta(x,y) X_h(\z;y)\\ X_h^\dagger(\z;x)\delta(x,y) & \rate_h(\z;x,y)
\end{bmatrix} 
 \succeq 0 
\end{align}
which for invertible $\rate_h(\z;x,y)$ is equivalent to the condition that $D_h^{00}(x,y)$ be positive semi-definite along with its Schur complement
\begin{equation}
D_h^{00}(x,y) - X_h^\dagger(\z;x)\rate_h^{-1}(\z;x,y) X_h(\z;y) \succeq 0
 \label{eq:positiveSchur}
\end{equation}
where $\rate_h^{-1}(\z;x,y)$ is the generalized inverse of $\rate_h(\z;x,y)$. The positivity condition for singular $\rate_h(\z;x,y)$ is a straightforward extension of this, see for example \cite{boyd2004convex}. We have since shown that the condition given by Equation \eqref{eq:positiveSchur}, holds for the zeroth, first and second moments of any consistent classical-quantum coupling, and provides an experimentally testable signature of classical gravity\cite{oppenheim2021gravitationally},

Turning back to Lorentz covariance, one might worry that in the master equation of Equation \eqref{eq:weinerfields}, there is a dependence on the choice $\partial_t$
\begin{align}
\partial_t\psiz=\mathcal{L}_t\psiz
\end{align}
As discussed in \cite{alicki-reldecoherence} we can consider any 4-vector $a$ from the future cone ${\cal F^+}$ i.e. $a\cdot a \geq 0$, and $a_0\geq0$, and consider the dynamics $\partial_a\psiz=\mathcal{L}_a$. Indeed, consider $\Lorentz$ an element of the proper orthochronous Lorentz group. Let $\pi(a,\Lorentz)$ be 
a representation of the Poincaré group and let the evolution of the state $\psiz$ be given by a completely positive trace-preserving Markovian map ${\cal E}_a$.
The Lorentz covariance condition as defined in \cite{alicki-reldecoherence} (see also \cite{holevo-weyl,poulinKITP}) is then
\beq
\pi(a,\Lorentz){\cal E}_b \pi(a,\Lorentz)^{-1}={\cal E}_{\Lorentz b}
\label{eq:covariant-me}
\eeq
for any $b \in {\cal F^+}$. In other words, if we perform a boost from the lab frame, evolve the state, then boost back into the lab, one gets the same final state as evolving the initial state in the lab for the boost time. The open quantum scalar field theory we will consider in the next section satisfies this property when the background space-time is flat, as can be seen from the fact that in that limit, it is equivalent to the Lindbladian considered in \cite{alicki-reldecoherence}. The relationship between this covariance condition, and Lorentz invariance, is discussed in \cite{UCLLorentz}.

\section{Post-quantum General Relativity}
\label{sec:PQgrav}

We now have a procedure to take a Hamiltonian which couples classical and quantum degrees of freedom, and use it to generate consistent hybrid dynamics. Let us now do that for general relativity coupled to quantum matter via the energy-momentum tensor. We will do so in the ADM formulation of general relativity\cite{arnowitt2008republication,dewitt1967quantum}, whose classical formulation we now recall.
One starts with a slicing up of space-time into space-like hypersurfaces labelled by the parameter $t$, and with coordinates $x$. The hypersurfaces are endowed with a 3-metric $g_{ab}$ with determinant $g$ at each point $x$. We will often drop the explicit $x$ dependence from field degrees of freedom, since all quantities should be assumed to be locally defined, unless they are a functional in which case this will be indicated with square brackets. We will put in the explicit $x$ dependence when we have two integrals over space, or if there could be any ambiguity. The components of the 3-metric are the canonical coordinates,
and $\pi^{ab}$ their conjugate momenta
	\begin{align}
	\pi^{a b}=\sqrt{g}\left(K^{a b}-g^{a b} K\right)
	\label{eq:pidef}
	\end{align}  
with	$K^{ab}$ the extrinsic curvature and $K$ its trace.
	
	 In general relativity, one is free to choose the time-like direction, by arbitrarily specifying a lapse $N$ and shift vector $N^a$ at each point $x$. The evolution of the hypersurfaces is then governed by the classical ADM Hamiltonian
\begin{align}
H_{ADM}[N,\vec{N}]:=\int dx \left(N\superhamtot+N^a\supermomtot_a\right)
+\oint ds\cdot \mathcal{H}^{(boundary)}
\label{eq:ADM}
\end{align}
and we now define the various terms appearing in it. $\superhamtot$ is called
 the total Superhamiltonian\foothide{The use of the term ``Superhamiltonian'' has nothing to do with the superhamiltonian used in the context of supersymmetry.\label{ft:super}} and is typically
 	written in terms of a pure gravity part, and a term which contains the matter fields
\begin{align}
\superhamtot=\superhamgrav
+\Tnnc
\label{eq:superham}\s.
\end{align}
 The first term is the purely gravitational part of the Superhamiltonian, given by
 \begin{align} 
 \superhamgrav:= G_{abcd}\pi^{ab}\pi^{cd}-\g^{1/2}R
 \end{align} 
with $R$ the intrinsic curvature on the 3-manifold analogous to a potential term, while the first term is analogous to a kinetic term, with $G_{abcd}$ the deWitt metric,
\begin{align}
G_{abcd}=\frac{1}{2}{g}^{-1/2}(g_{ac}g_{bd}+g_{ad}g_{bc}-g_{ab}g_{cd})\s .
\end{align}
 $\Tnnc$ %
 is the matter Hamiltonian density and %
 we will take it to also include the cosmological constant term $\sqrt{g}g_{ab}\Lambda_{cc}$.
 
 The total Supermomentum $\supermomtot_a$ is given by
 \begin{align}
 \supermomtot_a= %
 \supermomgrav_a+
 \Tnnc_a
 \label{eq:supermom}
 \end{align}
 with \phantomsection\label{par:deriv}
 $\superhamgrav_a:=-2\g_{ac}\D_b\pi^{cb}$ the purely gravitational contribution, and
 $\D_b$ the covariant derivative with respect to the Levi-Civita connection for the 3-metric. %
 $\Tnnc_a$ is the matter contribution.
 The final term in Equation \eqref{eq:ADM} is the boundary term,
 and we will ignore it in the present paper, although the 
 formalism can be extended to apply to its inclusion. Finally, it will prove convenient to define the total gravitational hamiltonian $\gravham\lapsh$ and the total matter hamiltonian $\matterham\lapsh$
 	as functionals so that  $H_{ADM}\lapsh=\gravham[N,\vec{N}]+	\matterham[N,\vec{N}]$ and
 \begin{align}
 \gravham[N,\vec{N}]:=\int dx \left(N\superhamgrav+N^a\supermomgrav_a\right),
 \s\s\s
 \matterham[N,\vec{N}]:=\int dx \left(N\ham+N^a\mom_a\right)
 \end{align}

With the Hamiltonian now defined let us turn to the equations of motion.
The dynamics is given by Hamilton's equations
\begin{align}
\frac{\partial \g_{ab}}{\partial t}=\frac{\delta H_{ADM}[N,\vec{N}]}{\delta\pi^{ab}}
, \,\,\,\,\, 
\frac{\partial \pi^{ab}}{\partial t}=
-\frac{\delta H_{ADM}[N,\vec{N}]}{\delta\g_{ab}}
\label{eq:gdotpidot}
\end{align}
where $\delta$ indicates
the Fr\'{e}chet derivative. 
In order for these dynamics to be invariant under the choice of the lapse $N$ and shift $N^a$, we need to impose primary constraints on their conjugate momenta
\begin{align}
P_N\approx 0, \,\,\,\,\, P_{N_a}\approx 0
\label{eq:p-constraints}
\end{align}
where $\approx$ denotes that the constraints are {\it weakly zero} in the sense that they 
only vanish on part of the phase space (the constraint surface).
To ensure that these primary constraints are conserved in time, i.e. that 
\begin{align}
\dot{P}_N=0, \,\,\,\,\,\dot{P}_{N_a}= 0
\end{align}
we require secondary
constraints 
\begin{align}
\superhamtot\approx 0, \,\,\,\,\, \supermomtot_a\approx 0
\label{eq:super-constraints}
\end{align}
The constraints, once initially satisfied, are preserved by the dynamical equations of motion.

One can write a Liouville equation for classical general relativity
\begin{align}
\frac{\partial\rho(g,\pi;t)}{\partial t}=
&\{H_{ADM}[N,\vec{N}],\rho\}
\nonumber\\
=&\{\gravham\lapsh,\rho\PBg+\{\matterham\lapsh,\rho\PBm+\{\matterham\lapsh,\rho\PBg
\label{eq:GRLiouville}
\end{align}
where we have divided up the Poisson bracket into one \jono{with respect to the pure gravitational degrees of freedom
\begin{align}
\{\matterham\lapsh,\rho\PBg&:=\int dx 
\Big(\frac{\delta\matterham\lapsh}{\delta g_{ab}}\frac{\delta \rho}{\delta \pi^{ab}}
-
\frac{\delta \matterham\lapsh}{\delta \pi^{ab}}\frac{\delta \rho}{\delta \g_{ab}}\Big)
\label{eq:GRPB}
\end{align}
and one with respect to the matter degrees of freedom, which for a scalar field with conjugate variables $\phi$,$\pi_\phi$ at each point $x$ would be
\begin{align}
\{\matterham\lapsh,\rho\PBm&=\int dx 
\Big(\frac{\delta\matterham\lapsh}{\delta \phi}\frac{\delta \rho}{\delta \pi_\phi}
-
\frac{\delta \matterham\lapsh}{\delta \pi_\phi}\frac{\delta \rho}{\delta \phi}\Big)\s .
\label{eq:matterPB}
\end{align}
The Poisson bracket with respect to the lapse and shift and their conjugate momenta is zero when the constraints are satisfied and is usually not included in the definition of the Poisson bracket. When we discuss the classical quantum theory, it may be useful to also include them in the definition of the Poisson bracket.}

When only matter is quantum, the matter Poisson bracket of Equation \eqref{eq:matterPB} just becomes the commutator $-i[\qmatterham\lapsh,\cqstate]$ which is unproblematic, and just represents the evolution of the quantum fields on a spatial manifold endowed with an evolving classical metric. This is just quantum field theory in curved space. However, the Poisson bracket of Equation \eqref{eq:GRPB} is the back-reaction of the quantum fields on space-time, and this is the dynamics we now seek to describe in terms of CQ-dynamics. 

We first replace $T^{\mu\nu}$ with the stress tensor of quantum field theory, $\qham=\sqrt{g}\T^{00}$. We can now write
\begin{align}
\qsuperhamtot=
\superhamgrav+
\qham
\end{align}
as long as we recall that this object is hybrid and cannot be used to derive classical or quantum evolution in the traditional way.
For the discrete master equation, we expand $\qsuperhamtot\xd$
in terms of local field operators $\L_\alpha(x)$, so that dropping the explicit $x$ dependence
\begin{align}
\qsuperhamtot\xd=\superhamgrav\id+h\ab\gpi \lbnox\lanox
\label{eq:cqlapse}
\end{align}
with $\L_0(x)=\id$ multiplying the purely classical part of the Superhamiltonian. For convenience, we explicitly separate out the pure gravity contribution $\superhamgrav$, and one can then take $h^{00}=0$. 
 Although in our derivation of the master equation, we took the $\L_{\alpha}$ to be a basis, and $\L_\ag$ to be traceless, one can show as with the Lindblad equation, that any set of Lindblad operators can be put into this standard form\cite{UCLPawula}, and so we will henceforth not assume that the other $\lanox$ are traceless.
We often consider matter which is {\it minimally coupled}, meaning $\qham$ doesn't depend on $\pi^{ij}$ so that we can write $h\ab(g)$. 
	For the continuous master equation, which we will consider elsewhere\cite{oppenheim2021constraints}, we use Lindblad operators $\L_{ij}\xd=\delta \qham\xd[N,\vec{N}]/\delta g_{ij}$, $\qham[N,\vec{N}]\xd$ and $L_0=\id$.

The Supermomentum can be treated similarly, where the stress-energy term in equation \eqref{eq:supermom} is replaced by
its quantum counterpart 
\begin{align}
\mom_a\rightarrow\qmom_a
\end{align}
leading to
\begin{align}
\qsupermomtot_a=\supermomgrav_a\id+p_a\ab\gnox\lbnox\lanox
\label{eq:cqshift}
\end{align} 
for the discrete master equation. Although the matrix with elements $N^ap_a\ab$ is not generally a positive semi-definite matrix, the combination $Nh\ab+N^ap_a\ab$ is.

To make things concrete it is useful to have in mind a scalar field, especially since the general case can be treated similarly and Lindbladian evolution of $\phi^4$ theory has been shown to be renormalizable in this case\cite{baidya2017renormalization}\phantomsection\label{par:caveats} albeit with the caveats mentioned in Section\ref{ft:renorm}. %
We have\cite{smear_foot}
\begin{align}
\qham
&=\frac{1}{2}\Big({g}^{-1/2}\pi_\phi^2+\sqrt{g}g^{ab}\nabla_a\phi\nabla_b\phi+\sqrt{g}m^2\phi(x)^2
-4\sqrt{g}\Lambda_{cc}\Big)
\label{eq:scalarTNN}\\
\qmom_a
&=\pi_\phi\nabla_a\phi
\end{align}
For the discrete evolution, we can now decompose the Hamiltonian as in Equation \eqref{eq:cqlapse}, with the metric degrees of freedom included in
$h\ab\gpi$. 
\begin{align}
h^{\pi\pi}:= 
g^{-1/2}, \s
h^{\phi\phi}:= 
\gfdet,\s
h^{ab}:=\gfdet g^{ab}
,\s
h^{\Lambda\Lambda}=\sqrt{g}
,\s
\label{eq:hlapse}
\end{align} 
with the rest zero, leaving the local Lindblad operators as
\begin{align}
\L_\pi\xd=\frac{1}{\sqrt{2}}\pi_\phi\xd,\s
\L_\phi\xd=\frac{m}{\sqrt{2}}\phi\xd, 
\s \L_a\xd=\frac{1}{\sqrt{2}}\partial_a\phi\xd
,\s %
\L_{\Lambda}={\sqrt{2\Lambda}}\proj{\Lambda}
\label{eq:scalarL}
\end{align}
Note that the $h\ab\gpi$ do not include the $N$ pre-factor, since we will be using it to apply the constraint equation, however, when we look at the dynamical equations of motion we will need to include it. We have chosen a simple form for $\L_{\Lambda}$, but {\it a priori}, other choices are possible, for example, the shift operator.
Their particular dependence is not important in terms of the broad outline of this discussion, rather, what is important is that we can decompose 
$\qham\xd$
in terms of local field operators. The expansion of the Supermomentum in terms of local operators could be treated similarly, 
with 
\begin{align}
\L_{\pi}\xd=\pi_\phi\xd,\s \L_{a}\xd=\partial_a\phi\xd
,\s \L_\id=\id\s.
\end{align}
Note that $\qmom_a\xd$ has no metric dependence and thus does not contribute to the back-reaction.%

Although we cannot generally define Fock states locally on a Cauchy slice, if we know for example that the dynamics will lead to a space-time which asymptotically has a time-like killing vector, then we could choose a vacuum state and
	\phantomsection\label{par:fock} 
set of creation and annihilation operators $\ann{p}(g)$,$\adag{p}(g)$ which depend on the metric to expand the matter Hamiltonian as
\begin{align}
\matterham\lapsh
= \frac{1}{2}\int \frac{d^3p}{(2\pi)^3}\omega_{\vec{p}}\Big( \adag{p}(g)\ann{p}(g)+\ann{p}(g)\adag{p}(g)\Big)
\label{eq:bbdag-decomp}
\end{align}
where %
we have dropped the 
cosmological constant term for simplicity. In a fixed background for example, the Lindblad equation using the Lindblad operators of Equation\eqref{eq:scalarL} are equivalent to those of Equation \eqref{eq:bbdag-decomp}. In this way, one sees the covariance of the master equation, however for general dynamical space-times an expansion of local operators in terms of creation and annihilation operators is not possible and we will use local Lindblad operators such as those
 of Equation \eqref{eq:scalarL}. %

The dynamical equations of motion are generated by the CQ-ADM Hamiltonian
\begin{align}
\cqadm[N,\vec{N}]=\int dx \big(N\qsuperhamtot+N^a\qsupermomtot_a\big)
\label{eq:CQADM}
\end{align}
with $\qsuperhamtot$ and $\qsupermomtot_a$ given by Equations \eqref{eq:cqlapse} and \eqref{eq:cqshift}. If we wish the dynamics to be invariant under the choice of lapse and shift, this would place strong constraints on it. 
It seems to
require that the stochastic part of the master equation, be generated by $N\rate_h\ab(\z|\z-\dist;x,y)$. I.e. that the model be linear in the sense that
$\rate\ab_{Nh}(\z|\z-\dist;x,y)$
can be replaced by
$N\rate\ab_{h}(\z|\z-\dist;x,y)$
and
$\rate\ab_{N_ap^a}(\z|\z-\dist;x,y)$
by $N_a\rate\ab_{p^a}(\z|\z-\dist;x,y)$. This is analogous to replacing $\{N\superhamtot,\rho\}$ by $N\{\superhamtot,\rho\}$ in the classical case, since they are equal when the constraint $\superhamtot\approx 0$ is satisfied.
	 In what follows, we take the lapse and shift outside the functional derivatives under the assumption that they
	either have no explicit metric dependence, or the constraints 
	are satisfied. The constraints in this theory will not be those of GR, but must be modified. As noted in \cite{UCL2022constraints}, it is not strictly necessary that the dynamics be invariant under the choice of $N$ and $N_a$, and there are sensible theories which break this such as shape-dynamics\cite{mercati2017shape, Anderson_2005, Gomes_2012}, or unimodular gravity\cite{einstein1952gravitational,van1982exchange,weinberg1989cosmological,unruh1989unimodular,alvarez2005can,smolin2009quantization, shaposhnikov2009scale}, but it is desirable to have some local symmetry, even if it isn't full diffeomorphism invariance.\label{ft:lapse_linear}.
	
We will not restrict ourself to a particular realisation of $\rateabxd$ and $\rate_{p_a}\ab(\z|\z-\dist;x,y)$ at this point, but we will assume that the realisation satisfies this linear condition and then present realisations which have this property.
We thus have the master equation
\begin{align}
\frac{\partial\varrho(\z;t)}{\partial t}=
&\int dxN\{\superhamgrav,\psiz\}
+\int dx N^a\{\supermomgrav_a,\psiz\}
-i[\matterham\lapsh,\psiz]
\nonumber\\
&+\int dx dy \ddf \Big[N\rate_{h}\ab(\z|\z-\dist;x,y)\lax\cqstate(\z-\dist)\lby
-\frac{1}{2}N
\rate\ab_h(\z;x,y)\{\lby\lax,\psiz\}_+\Big]
\nonumber\\
&+\int dxdy \ddf \Big[N_a\rate_{p^a}\ab(\z|\z-\dist;x,y)\lax\cqstate(\z-\dist)\lby-
\frac{1}{2}
N_a\rate\ab_{p^a}(\z;x,y)
\{\lby\lax,\psiz\}_+\Big]
\label{eq:dynamicalPQG}
\end{align}
where we write $\z$ for the phase space degrees of freedom when $g_{ab}$,$\pi^{ab}$ are too cumbersome and have absorbed the  bare term $\linrate\ab(\z;x,y)$ into $\rate\ab_{h}(\z|\z-\dist;x,y)$. The functional integral $\ddf$ is over the phase space. We have here taken the lapse $N$ and shift $N^a$ outside the Poisson bracket, under the assumption that the constraints (suitably modified from the deterministic theory) are satisfied\cite{UCL2022constraints}.

The first two terms of Equation \eqref{eq:dynamicalPQG}
corresponds to the deterministic classical evolution associated with the Lindblad operators $\L_0\xd=\id$. %
The dynamics associated with this
classical part could be stochastic, and could {\it a priori} include jumping or diffusion terms as well as friction terms. We take such terms to be included in $\rate_{h}^{00}(\z|\z-\dist;x)$.
Because they are associated with $\L_0$ they do not need to be positive.

The next %
term is the commutator of the state with the matter Hamiltonian %
describing how quantum matter evolves in a background space-time. Finally, we have the interaction terms between gravity and matter which must be stochastic. Two particular choices of $\rate\ab(\z;x,y)$ are given as Equations \eqref{eq:modelcov} and \eqref{eq:dynamicsPQG-continuous} and in both cases, the integral over $\dist$ can be removed. \phantomsection\label{par:in_over_dist}%
  As shorthand, we will denote Equation \eqref{eq:dynamicalPQG} by 
\begin{align}
\frac{\partial\varrho\gpit}{\partial t}&= {\mathcal L_\cqadm}\varrho\gpit
\nonumber\\
&=\int dx\big[ N{\cqhamcon}\varrho\gpit+N^a{\mathcal L_a}\varrho\gpit\big]
\end{align}
Because %
$\qmom_a\xd$ has no metric dependence, the evolution along $N^a$ can be pure Lindbladian or purely deterministic. In \cite{UCL2022constraints}, we argue that diffeomorphism invariance requires 
the latter, i.e. that $\rate_{p^a}\ab(\z|\z-\dist;x,y)=\rate\ab_{p^a}(\z;x,y)=0$. 
Equation \eqref{eq:dynamicalPQG} then simplifies to 
\begin{align}
\frac{\partial\psiz}{\partial t}&
=\int dx\Big[ N\{\superhamgrav,\psiz\PB+N^a\{\supermomgrav_a,\psiz\PB\big]
-
i[\matterham\lapsh,\psiz]
\nonumber\\
&+\int dxdyN \ddf \Big[\rate_{h}\ab(\z|\z-\dist;x,y)\lax\psizd\lby
-\frac{1}{2}
\rate\ab_h(\z;x,y)\{\lby)\lax,\psiz\}_+\Big]
\label{eq:dynamicalPQGsimpler}
\end{align}

This should be thought of us the fundamental dynamical equation, but
we can perform a Kramers-Moyal expansion on it. Since we expect all moments to appear, the moment expansion will include terms such as
$\frac{\delta h\ab}{\delta g_{ab}}\frac{\delta\psiz}{\delta \pi^{ab}}$ since it transforms like a density. This moment at 
zeroth order in $\tau$ is needed to recover the classical limit of general
relativity. 
We now require a positive transition matrix $\rate_h\ab(\z|\z-\dist;x,y)$  (except for the $\rate_{h}^{00}(\z|\z-\dist;x,y)$ component), and that each moment transform as a density to ensure invariance of the dynamics under spatial diffeomorphism. An example of such a realisation for the discrete master equation is given in \cite{UCL2022constraints}, namely for $\alpha,\beta\neq 0$, 
\begin{align}
\int dy \ddf N\rate_h\ab(\z|\z-\dist;x,y)\lax\psizd\lby
=
\frac{N}{\tau}
\exp\Big[{\tau{({h^{-1})}(\z;x)}\{h(\z;x),\cdot\PB}\Big]^\alpha_\gamma h^{\beta\gamma}(\z;x)
\lax\psiz\lbx
\label{eq:modelcov}
\end{align}
where $\exp[M]^\alpha_\gamma$ are the components of the matrix exponential of the matrix $M$ and it is purely local in $x$. Here, $M$ has components $M_i^k=\Big[{\tau{({h^{-1})_{ij}}(\z)}\{h^{jk}(\z),\cdot\PB}\Big]$ and  $\rate\ab(\z;x,y)=\frac{1}{\tau}h\ab(\z;x)\delta(x,y)$. 
and gives a Kramers-Moyal expansion, up to constants of
\begin{align}
\int& \rate_h\ab(\z|\z-\dist)\psizd\ddf
=
\frac{1}{\tau}
h\ab(g_{ab})
\psiz
+
\frac{\delta h\ab(\z)}{\delta g_{ab}}\frac{\delta\psiz}{\delta \pi^{ab}}
-
\frac{\delta h\ab(\z)}{\delta \pi^{ab}}\frac{\delta\psiz}{\delta g_{ab}}
\nonumber\\
&
+
\tau D^{abcd,\alpha\beta}(\g)
\frac{\delta^2\psiz }{\delta \pi^{ab}\delta \pi^{cd}}
+\cdots
\label{eq:GRfirstfewmoment}
\end{align}
where the diffusion terms $D^{abcd,\alpha\beta}(\g)$, and all higher order moments are tensor densities of the appropriate weight to contract with the derivatives acting on the density matrix. 
The absence of functional derivatives with respect to $g$, higher than $\delta \varrho/\delta g_{ab}$ means that this is a {\it Brownian realisation}. As a result,
Equation \eqref{eq:pidef} is unmodified and we only have diffusion in $\pi^{ab}$. Higher order functional derivatives in $\pi^{ab}$ can be suitably smeared to control divergences. If we want to ensure that our evolution equation is invariant under canonical transformations, we can express it in terms of Poisson brackets, with the second order term being written for example, as $\{J_o,\{J_i,\psiz\}\}$. It may also be necessary to include the lapse, shift and their conjugate momenta in the Poisson bracket, which is a realisation we will pursued in \cite{oppenheim2021constraints}.

There is more freedom when it comes to $\rate_{h}^{00}(\z|\z-\dist;x,y)$ which could include dispersion terms, non-Brownian terms, or even friction terms such as those of the form
\begin{align}
\frac{\delta \gamma^{ab}\varrho}{\delta \pi^{ab}\xd}
,\,\,\,\, \gamma^{ab}\xd:=\eta^{abcd}\xd\frac{\delta\superhamgrav}{\delta \pi^{cd}\xd}
\end{align}
with $\eta^{abcd}$ a local function of the metric. %
However, gauge invariance places strong restrictions on these terms, and here we leave them unspecified and denote them by the differential operator $\Gamma^{00}\xd\varrho$.
The resulting evolution law is
\begin{align}
\frac{\partial\psiz}{\partial t}=
&
\int dx\Big[N \{\superhamgrav,\psiz\PB+N^a\{\supermomgrav_a,\psiz\PB\Big]
-
i[\matterham\lapsh,\psiz]
\nonumber\\
&+\int dx \frac{1}{\tau}
Nh\ab(\z)
\Big[ \lanox\psiz\lbnox
-
\frac{1}{2}\{\lbnox\lanox,\psiz\}_+\Big]
+\int dx N \Gamma^{00}\xd\psiz
\nonumber\\
&+\int dx N
\Big[
\frac{\delta h\ab}{\delta g_{ab}\xd}\lanox\frac{\delta\psiz}{\delta\pi^{ab}}\lbnox
+\tau D^{abcd,\alpha\beta}_{\pi}(\g) 
\frac{\delta^2 }{\delta \pi^{ab}\pi^{cd}}
+\cdots
\label{eq:dynamicalPQG-expanded}
\end{align}

Let us now see that in the classical limit, we recover general relativity to leading order. For convenience of presentation, we consider the case where the quantum system is uncorrelated with the classical system. The more general case is akin to Equation \eqref{eq:semiclassical-correlated-example}. Recalling Equation \eqref{eq:cqRef3}, that $\rho(\z;t):=\tr\psiz$, we perform the trace on the quantum system in Equation \eqref{eq:dynamicalPQG-expanded}. Dropping terms of second order and higher\phantomsection\label{par:trace}
 we are left with
\begin{align}
\frac{\partial\psiz}{\partial t}=
&\int dx N\{\superhamgrav\xd,\rho\PB+\int N^a dx\{\supermomgrav_a\xd,\rho\PB
+\int dx\sqrt{g}\tr N \{ \qham\xd,\varrho\PB
+\cdots
\label{eq:semiclassical-mastereqn_first_order}
\end{align}
which, when the constraints of GR are satisfied, is the classical Liouville equation for the gravitational degrees of freedom of general relativity sourced by quantum matter. The third term gives the back-reaction we expect. We thus see that in the classical limit we recover Einstein's equations. To leading order, the evolution of the quantum fields is seen from Equation \eqref{eq:dynamicalPQG-expanded} to correspond to the ordinary commutator terms (quantum field theory in curved space) plus a Lindbladian term.

A realisation which gives the continuous master equation will be discussed in \cite{oppenheim2021constraints} and \cite{oppenheim2021gravitationally} with  Šoda, Sparaciari and Weller-Davies, and is given by%
\begin{align}
\frac{\partial\cqstate}{\partial t} = &
\int dx N\{\H\xd,\cqstate\}+\int dx N^a\{\H_a\xd,\cqstate\}
 -i  \left[\qhamint[N,\vec{N}],\cqstate  \right]
 \nonumber\\
 &
+\frac{1}{2}\int dx N \left(\{\qham\xd,\cqstate)\}-\{\cqstate,\qham\xd\}\right)
+\frac{1}{2} \int 
dxdy 
N \lambda_{\mu\nu\hat{\sigma}\tau}(g;x,y) \left[\L^{\mu\nu}(x),\left[\cqstate,\L^{\hat{\sigma}\tau}(y)\right]\right]
\nonumber\\
&
+\frac{1}{2} \int 
dx dy 
N\{D_2\ab (x,y)J'_\alpha(g;x),\{J_\beta(g;y),\cqstate\}\}
\nonumber\\
\label{eq:dynamicsPQG-continuous}
\end{align}
when the constraints are satisfied. 
Here, $\L^{ij}\xd:= \frac{\delta \qham\xd}{\delta g_{ij}}$, $\L^{nn}\xd:=\qham\xd$,
and the pure Lindbladian couplings $\lambda_{\mu\nu\hat{\sigma}\tau}(g;x,y)$ consist of submatrices with elements
$\lambda_{ijkl}:=D_{0,ijkl}(g;x,y)$, $\lambda_{ijnn}(g;x,y)$, $\lambda_{nnkl}(g;x,y)$ and the constant $\lambda_{nnnn}(g;x,y)$. Here, Greek indices are akin to space-time indices $\mu=n,i$, and we use $n$ instead of $0$ since $0$ already corresponds to the $\id$ Lindblad operator.

The condition for complete positivity can be found from Equation \eqref{eq:positiveSchur}. Denoting the 
generalised inverse of the matrix $D_0(g;x,y)$ as $D_0^{-1}$, we require that the matrix with elements $D_0^{ijkl}(x,y)$ and
$D_2\ab\frac{\delta J'_\alpha(g;x)}{\delta g_{ij}}\frac{\delta J_\beta(g;y)}{\delta g_{kl}}-\left(D_0^{-1}\right)^{ijkl}(x,y)$ 
be positive semi-definite as a matrix and kernel to guarantee complete positivity of the master equation. $D_2\ab(x,y)$, $J_\alpha(N,\vec{N},g;x)$ and $D_{0,ijkl}(N,\vec{N},g;x,y)$ are chosen so that the realisation is linear in the lapse and shift. If we take the $J_\alpha$ to be local in the metric, the simplest case is then $J=\sqrt{g}$, so that the diffusion term is $\frac{1}{2}\int dxdy N(x)D_2(x,y)\sqrt{g(x)}\sqrt{g(y)}g^{ij}(x)g^{kl}(y)\frac{\delta\cqstate}{\delta \pi^{ij}(x)\delta\pi^{kl}(y)}+\cdots$, but there are other more complicated realisations which also recover the dynamical Einstein's equation as a limited case (by adding friction and additional diffusion terms for example). The additional Lindbladian couplings corresponding to the Lindblad operator $\qham$ are not needed in order to reproduce general relativity in the classical limit, nor to preserve positivity, but instead appear to be required by gauge invariance. The constraint equations for this realisation are derived in \cite{oppenheim2021constraints} by following the procedure outlined in \cite{UCL2022constraints}. Alternatively, one can include the lapse, shift, and their conjugate momenta as canonical degrees of freedom and thus part of the Poisson bracket in Equation \eqref{eq:dynamicsPQG-continuous}.

This realisation has the clear advantage of being continuous, while the former realisation has the advantage of being linear in the Hamiltonian. 
In terms of the evolution of the quantum system, for each classical space-time manifold, the
quantum system undergoes both decoherence, as well as unitary evolution generated by the Hamiltonian, as well as quantum jumps which are accompanied by a back-reaction on the space-time.
The diffusion terms imply that the solution of Equations \eqref{eq:dynamicalPQG-expanded} and \eqref{eq:dynamicsPQG-continuous} (or alternatively \eqref{eq:dynamicsPQG-continuous}) should be a distribution which has a variance about the solution to Einstein's equations 
	when we perform the trace over the quantum system. This presents a testable deviation from classical general relativity, potentially observable when the other terms are small, such as at low acceleration\cite{acceleration_foot}. The absence of such terms is likely to falsify this model. On the other hand
since diffusion of the metric can result in stronger gravitational fields when we might otherwise expect none to be present, it raises the possibility that diffusion may explain galaxy rotation curves\cite{rubin1970rotation} and galaxy formation without the need for dark matter.

In addition to the dynamical equation of motion, Equation \eqref{eq:dynamicalPQG} or \eqref{eq:dynamicalPQG-expanded}, we need to impose the constraints of general relativity. Let us recall how this is done classically. There, the Hamiltonian and momentum constraint of Equation \eqref{eq:super-constraints} follows from demanding that the Lagrangian be invariant under arbitrary time reparametrizations, and spatial diffeomorphisms. 
This leads to the primary constraints on their conjugate momenta, $p_N=0$ and $p_{N^i}=0$. 
Preservation of the primary constraints is implemented in the ADM formalism by requiring that $\dot{p_N}=0$ and $\dot{p_{N^a}}=0$ (the Supermomentum and Superhamiltonian constraints). These are just constraints on the classical phase space, and so we can constrain $\rho$ to lie on the surface in phase space where the constraints 
are zero. 

Were we to impose the constraint in an equivalent manner here,  it would amount to demanding that $\cqadm$ be invariant under the action of $P_N$ and $P_{N_a}$. This would lead to a Hamiltonian constraint
of the form
\begin{align}
\superhamgrav+16\pi G\qham\stackrel{?}{\approx} 0
\label{eq:badconstraint}
\end{align}
at every $x$ which would be impossible to satisfy, since it contains both a classical term $\superhamgrav$ and a local quantum operator $\qham$.

However, such a constraint is not needed -- rather, we should instead ask, what restrictions are required such that the equations of motion satisfy the required symmetry.
In the classical case, we demand that physical degrees of freedom do not depend on the choice of lapse $N$ and shift $N^a$ i.e. that these are gauge choices. This requirement
necessarily leads to the constraints of general relativity. Likewise here, we should not first demand that the constraints be satisfied, but rather, require that the equations of motion be invariant under the  choice of lapse and shift. This turns out to lead to modified constraints which are not of the form
\eqref{eq:badconstraint}. We discuss how the constraints can be generated from gauge invariance of the equations of motion in \cite{UCL2022constraints}.  Invariance of the equations of motion under spatial diffeomorphisms is already guaranteed as the evolution equation involves only integrals over scalar densities. 
 
In order to find a solution to the dynamical and constraint equations, we need the equations to be compatible. Attempts to quantise gravity have floundered on this point, as the constraints are operators and typically don't commute, so no simultaneous solution of the constraints exist. 
This is the case in loop quantum gravity, although it is hoped that one can find an operator ordering such that the constraints close. Likewise string theory is background dependent, which presumably means that in the full theory, gauge invariance is broken. 
While it is hoped that frameworks for String Field Theory could result in a background independent approach, it's not clear whether this can be maintained. 
Here, we also require constraint compatibility, but it takes on a different form. Because we have attempted to implement gauge invariance on the level of the equations of motion rather than as operator identities, the relevant notion of constraint compatibility is not simply the commutation of operators, but whether the evolution operators and constraints and generators "commute" with one another when applied to the state. In other words, we require for example
\begin{align}
\cqmomcon(x)\cqhamcon(y)\psiz-
\cqhamcon(y)\cqmomcon(x)\psiz
\approx 0
\label{eq:compat}
\end{align}
and similarly we require that the "commutation relations" for all other generators vanish on the constraint surface -- the right-hand side must be a linear combination of generators and constraint equations. This notion of constraint compatibility replaces the closure of the Dirac algebra in the classical theory. 

This program is initiated in \cite{UCL2022constraints} in the Heisenberg representation to find gauge invariant observables $A$.
This constraint structure, while involved, is far simpler than the constraint structure of loop quantum gravity or canonical quantisation, and provides a model in which to understand some of the conceptual challenges these theories face.
While the constraints are more complicated because they are implemented on the level of equations of motion, the classical nature of the gravitational field means they don't suffer from operator ordering issues and are more tractable. Nonetheless, it may be  impossible to find a realisation of the theory, and a symmetry principle, such that the set of constraint equations close. In that case, one could still impose the standard constraints of general relativity as initial conditions, as is done in the ADM formulation, but they may not be conserved in time. One could instead, impose the standard constraints as expectation values, but again, it's possible these may also not be conserved in time. This would likely mean the theory was not invariant under diffeomorphisms of the $3+1$ dimensional space-time, but it would not render the theory inconsistent, since we have here formulated the theory, not as a $3+1$ theory of space-time, but rather, 
 as a Hamiltonian theory of the spatial metric. The evolution of matter is governed by the Hamiltonian and Lindblad operators, which are functionals only of the three metric. To be consistent, the theory needs to be invariant under spatial diffeomorphisms, which it manifestly is. However, although it would be a consistent theory, it would be unclear how to justify imposing the constraint as a mere initial condition. So while the theory may be consistent, it would require further motivation. 
 
 An alternative approach can  be pursued, now that classical-quantum path integral methods have recently been developed\cite{oppenheim2023path}.  Since the path integral is computed in configuration space, this allows for the construction of manifestly diffeomorphism invariant theories. These have since been proposed together with Weller-Davies\cite{oppenheim2023covariant}. One, based on the trace of Einstein's equations, is different to Einstein's general relativity, but does at least demonstrate that diffeomorphism invariant classical-quantum theories are possible in principle. A theory which contains all components of the Einstein equations is also proposed but it is not yet known whether dynamics which deviate from general relativity are suitably suppressed.

\section{Discussion}
\label{sec:discussion}

The general form of hybrid classical-quantum dynamics has been presented here as Equation \eqref{eq:cq-dynamics}, which includes discrete evolution, and a continuous form\cite{UCLPawula}. Based on this, we have here exhibited a theory of classical general relativity coupled to quantum field theory which is invariant under spatial diffeomorphisms. The theory is consistent -- the dynamics is completely positive, norm preserving, and linear in the density matrix. We have here specified the dynamics up to a choice of the positive semi-definite matrix $\rate\ab_{h}(\z|\z-\dist;x,y)$.
The choice is constrained by gauge invariance, and thus must satisfy further criteria, as discussed in \cite{UCL2022constraints}. Two particular choices of $\rate\ab_{h}(\z|\z-\dist;x,y)$, are the model of Equation \eqref{eq:modelcov} and that corresponding to Equation \eqref{eq:dynamicsPQG-continuous}. These are further considered in \cite{UCL2022constraints} and \cite{oppenheim2021gravitationally,oppenheim2021constraints} respectively along with the constraint equations which correspond to them.
In \cite{UCLPawula}, it is shown that the first realisation leads to discrete jumps in phase space, while the second realisation follows from demanding that the dynamics be continuous in phase space.  The first moment in their Kramers-Moyal expansion corresponds to general relativity in the ADM formulation, and thus both recover classical general relativity as a limiting case. \phantomsection\label{par:realisations}  A proposal based on the path integral has been introduced in \cite{oppenheim2023covariant}.

We required that\phantomsection\label{par:require} the metric remains classical even when back-reacted upon by quantum fields, and this has two important consequences. First, the interaction is necessarily stochastic, which is important, since a principle motivation behind the theory were attempts to resolve the black-hole information problem.
That there exists a fully quantum theory of gravity has been an argument against theories which have fundamental information loss, in the sense that given any such dynamics one can always find a purification of the state onto some environment, such that the full theory is unitary. 
In this case,  Coleman has shown that when we trace out the environment, the dynamics is equivalent to a unitary theory with unknown coupling constants\cite{coleman1988black}. 
This renders the information loss trivial: in a universe with a finite number of random coupling constants, an observer only has to perform a sufficient number of experiments to determine these coupling constants, after which time the theory will behave arbitrarily close to a unitary one from the  point of view of this experienced experimentalist.\phantomsection\label{par:coleman} However, the argument in \cite{coleman1988black} need not apply here. Any purification which keeps some degrees of freedom classical would need to obey an unnatural condition, namely that it remain of the form
\begin{align}
\varrho_{AB}=\sum
\ket{g,\pi}_A
\otimes\ket{g,\pi}_B
\otimes\ket{\varrho_{matter}(g,\pi)}_{AB}
\label{eq:TFD}
\end{align}
with system $A$ encoding the
classical degrees of freedom and system $B$ being the purifying system
\cite{TFD_foot}. For these degrees of freedom to remain classical, one further needs to impose the condition that one can only measure in the basis where $\ket{g,\pi}_A$ is diagonal. If one lifts this restriction on measurements, or allows dynamics more general than the one preserving the form of the state of Equation \eqref{eq:TFD}, then it is far from clear that the dynamics would still be completely positive and trace-preserving. The dynamics here only needs to give positive probabilities on measurements which commute with the basis representing the classical degrees of freedom. It does not need to give positive probabilities for all measurements. Thus we cannot conclude that this theory has consistent dynamics on the purified state. Or to put it another way, it is unlikely to be the case, that the dynamics considered here cand be viewed as unitary quantum dynamics with a local environment which is traced out. The fact that the quantum system couples to a classical one, thus provides a potential loophole to the no-go theorem of \cite{coleman1988black} ruling out non-trivial information destruction, while as argued in the introduction, information destruction provides the loophole out of the no-go argument of Feynman and Aharonov against quantum systems interacting with classical ones. The two no-go theorems annihilate each other.

\phantomsection\label{par:tension}
	There may appear to be some tension  between the claim that the theory may evade the no-go argument of Coleman, and the idea that the theory presented here could be an effective theory which emerges as the classical limit of a fully quantum theory. However, this is not necessarily in contradiction.
	In \cite{UCLQQtoCQ}, we consider fully quantum descriptions, which effectively act like a classical-quantum description. The classical system corresponds to a quantum system decohered into a family of coherent states, and for some potentials, this desciption is valid for long times. Since this description is fully quantum, Coleman's argument would imply that it corresponds to a unitary theory on an enlarged Hilbert space with unknown coupling constants, and that there exist measurements which would allow an experimenter to determine these constants.  However, in the classical-quantum description of this fully quantum theory, we do not allow arbitrary measurements on the ``classical'' system -- only measurements of $q$ and $p$ are allowed. Coleman's argument allows the experimenter to perform measurements in an arbitrary basis and it appears unlikely one could use fixed measurements to determine the coupling constants.

The second consequence of having a quantum-classical interaction is that the measurement postulate of quantum theory is no longer needed. One has fundamental decoherence of the quantum field by the classical metric. In this view, classicality is not an emergent or effective property of large quantum systems, but rather, fundamental.
In the discrete theory discussed here, $h\ab$ is diagonal and unique for each Lindblad operator. As with the Stern-Gerlach example, the accompanying jump in phase space unambiguously determines which Lindblad operator was applied to the quantum state. If we know the initial pure state of the quantum system, then by monitoring the classical system, we know which sequence of Lindblad operators were applied to the quantum state and at what times. As such, the quantum state conditioned on the classical degrees of freedom remains pure. Likewise, in the continuous case, we have since found that 
there is a very natural class of theories in which the quantum state remains pure conditioned on the classical trajectory\cite{layton2022semi}.
	This is 
ironic given the motivation of obtaining dynamics which allowed for the destruction of information in black-hole evaporation. The dynamics here, while stochastic, can leave the quantum state pure. It is the classical degrees of freedom which gain entropy.
This suggests that in black-hole evaporation, the associated entropy is the entropy of the classical space-time. Since the formalism introduced here allows one to study the back reaction of matter fields on the gravitational field, there is hope that we can better understand black-holes, as well as other systems where gravity and quantum effects are important.

Let us next address a number of open questions and challenges that the theory faces, starting with the objection to theories of information destruction due to Banks, Peskin and Susskind\phantomsection\label{par:bps} (BPS)\cite{bps}. BPS argued
that a local interaction which leads to decoherence will 
lead to anomolous heating to such a degree that it will contradict observation. Although the BPS argument was not made in the context of classical-quantum theories, it does apply to Lindbladian evolution of the quantum state $\hat{\sigma}$ with Lindblad operators being local field operators, and our theory does contain such terms. In particular, BPS considered an evolution equation
\begin{align}
\frac{\partial\hat{\sigma}(t)}{\partial t}
&=-i[\qmatterham,\hat{\sigma}(t)]
+ \int dxdx'
D_0\ab(x-x')\left( 
\L_{\alpha}(x)\hat{\sigma}(t)\L_{\beta}^\dagger(x')
-
\frac{1}{2}\{\lbxp\lax,\hat{\sigma}(t)\}_+ \right)  
\end{align}
 and couplings $D_0\ab(x-x')$.
The Hamiltonian of the matter fields $\qmatterham$ will not be conserved, and when $D_0\ab(x-x')$ has short range (decays rapidly with $|x-x'|$) there may be significant anomalous heating. In essence, local operators $\L_{\alpha}(x)$ don't commute with the Hamiltonian, and in the limit where $D_0\ab(x-x')$ is a delta function, the average rate of energy production contains terms which diverge. This is easily seen in the Heisenberg representation, since when $D_0\ab(x-x')$ is symmetric in $x,x'$, and $\alpha,\beta$ 
\begin{align}
\frac{d{\qmatterham}}{dt} =\frac{1}{2}\int dx dx' D_0(x-x')[\L_{\alpha}(x),[\qmatterham,\L^\dagger_\beta(x')]]
\label{eq:heating}
\end{align} 
If the $\L_{\alpha}(x)$ contain only a finite number of derivatives of local field operators, then the commutator gives terms proportional to $\delta(x-x')$, or $k$-th derivatives $\nabla^k\delta(x-x')$. These term will diverge if $D_0(\epsilon)$ diverges as $\epsilon\rightarrow 0$. For example, if $D_0(x-x')\propto \delta(x-x')$. 

This should trouble us if it contradicts observation, but need not trouble us on a fundamental level. If the dynamics is Lindbladian, there is no reason to expect energy conservation -- the dynamics isn't unitary and so Noether's theorem doesn't apply. The generator of time-translations is not the Hamiltonian but the Lindbladian\cite{Noether_foot}. In fact, there is no way to canonically define energy, since by choosing a different basis of Lindblad operators, the Hamiltonian will change.
In the case of classical-quantum systems, the notion of energy becomes even harder to  define, since the Hamiltonian is a hybrid object. 
On top of this, in general relativity there is no local definition of energy, nor conservation law for energy integrated over an infinitesimal spatial surface\cite{brown1992quasilocal}.  From a physical point of view, this is arguably the more relevant quantity, and so the question of energy conservation is not a matter of principle, but rather, consistency with experiment. 
One must ensure that the vacuum is stable enough that it doesn't contradict experimental observation.

The vacuum can be made stable, by turning down the Lindbladian coupling so that $D_0(\epsilon)$ descreases as $\epsilon\rightarrow 0$. Since $d{\qmatterham}/dt$ goes as $D_0(\epsilon)$, we can regularise it by for discretising space-time, and scaling $D_0(\epsilon)$ with $\epsilon$. 
In the pure Lindbladian case, there is no reason not to take the Lindbladian coupling to be arbitrarily small, in which case the effect is weak in flat space and only strong when the 3-metric becomes singular, since we expect $D_0(x-x')$ to contain terms proportional to $\sqrt{g(x)}$ and $\sqrt{g(x')}$. Seen in this light, the problem raised by BPS is merely one of having to introduce another scale.

A similar issue occurs in spontaneous collapse models\cite{ballentine1991failure,gallis1991comparison,shimony1990desiderata}, but in and of itself, this does not present a problem, since again, one can tune down the Lindbladian coupling $D_0(\epsilon)$ as small as one wishes. Those who advocate for spontaneous collapse models typically require $D_0(x-x')$ to be large enough to account for the level of decoherence we see around us. This does not need to be the case here where environmental decoherence is expected to account for most of the decoherence we observe, and where the ``collapse'' of the wave-function is a by-product which only occurs when sufficient correlations has build up between the gravitational field and the quantum state together with its correlated environment. Furthermore, decreasing $D_0(\epsilon)$ requires more gravitational diffusion\cite{oppenheim2021gravitationally} which causes additional decoherence once the gravitational field is integrated out\cite{tilloy2016sourcing,UCLcoherence}. The heating effect of this secondary decoherence is a source of concern, but appears to be milder than in BPS\cite{UCLcoherence,oppenheim2021gravitationally}, although whether this is within experimental limits requires further investigation since at short distances, we are operating in a regime where relativistic effects become important.

The question of whether the couplings can be tuned down as we decrease $\epsilon$ is related to the question of whether the theory is regularisable\cite{oppenheim2023path}. 
In the case of a scalar field with a single Lindblad operator $\phi(x)$, the Lindbladian dynamics is both renormalisable and Lorentz invariant\cite{baidya2017renormalization,UCLLorentz}, and we can choose the coupling constant to generate as little anomalous heating as we wish, since we do not require the amount of entropy production to be large.
 Understanding whether the theory can be regularised, remains an important open question for the program presented here. The question is complicated by the fact that relativistic corrections in the gravitational field become important at short distances\cite{UCLDMDNE}, and by the fact that some aspects of the heating appear to be gauge artefacts\cite{layton2023weak}. 

Other strategies have been considered, but they may make the theory less compelling. One may be able to make the anomalous heating small by introducing a small violation of locality as has been attempted in the case of Lindbladians in the context of field theories\cite{poulinKITP}, or as is done in spontaneous collapse models where the coupling is modified at a length scale at the order of the proton radius. 
It is worth emphasising that if one reduces heating by allowing  $D_0\ab(x-x')$ to have finite range, this doesn't necessarily violate locality. If $\L_{\alpha}(x)$ are local operators, then it's possible for local observables to still evolve locally even though $D_0\ab(x-x')$.  The price we pay if $D_0\ab(x-x')$ doesn't fall off fast enough with  $|x-x'|$, is that correlations can be created over finite distances\cite{bps, OR-intrinsic}. While this seems to suggest some  correlated environment should be evoked, it doesn't lead to superluminal signalling or inconsistencies. It will however naively violate Lorentz invariance, since one has spatial correlations but no temporal ones. This could be rectified by taking $D_0\ab(x-x')$ to not  depend on the spatial distance $|x-x'|$, but rather the space-time distance. For example, one could take it to be the Green's function of the Laplace-Beltrami operator. This would would then suggest a non-Markovian evolution, reminiscent of an effective theory rather than a fundamental one. 
Another alternative is to modify gravity at short distances. Since gravity is only tested to within the milometer scale, such modifications have often been considered, but would likely necessitate some additional structure such as extra dimensions\cite{arkani1998hierarchy}.

Whether such measures are required is unclear. In the context of the theories considered here, the situation is significantly different from the pure Lindbladian evolution considered by BPS. First, the action of the Lindblad operators on the quantum field is accompanied by a back action on the classical degrees of freedom, and in particular, the gravitational degrees of freedom. While the argument of BPS suggests there will be an increase in the expectation value of the energy density of the matter Hamiltonian density $\qham$, this can be accompanied by a decrease in the energy of the classical system due to this back-reaction. %
In this case, this is the  kinetic energy density $G_{ijkl}\pi^{ij}\pi^{kl}$ of the pure gravity part of the Superhamiltonian, $\superhamgrav$. In the case of a discrete realisation of this theory, such as Equation \eqref{eq:modelcov}, the action of the Lindblad operator is accompanied by a jump in $\pi^{ij}(x)$. Likewise, the diffusion term of the continuous realisation, Equation \eqref{eq:dynamicsPQG-continuous} acts on the kinetic term of $\superhamgrav(x)$ in the Heisenberg representation producing an energy density flux of $D_2^{ijkl}(x,x)G_{ijkl}(x)$ \phantomsection\label{par:negflux} which can be negative and state independent, and also needs to be regularised. For example, consider
\begin{align}
\lim_{x-x'\rightarrow \ell}D_2^{ijkl}(x,x')=D_2
\frac{\sqrt{g}N}{2}\left(g^{ik}g^{jl}+g^{il}g^{jk}-2\beta g^{ij}g^{kl}\right)
\label{eq:diffusion_co}
\end{align} 
with $D_2$ some constant which depends on the cut-off $\ell$, and $\beta<1/3$ to ensure $D_2^{ijkl}$ is positive semi-definite. If we wish $D_2^{ijkl}(x,x')$ to only include the final
	term in Equation \eqref{eq:diffusion_co} proportional to $g^{ij}g^{kl}$, then we can take $D_2\rightarrow D_2/\beta$ and $\beta\rightarrow-\infty$.

Using the diffusion coefficient of Equation \eqref{eq:diffusion_co} we can compute the apparent flux of gravitational kinetic energy density using the continuous realisation of Equation \eqref{eq:dynamicsPQG-continuous}  
in the Heisenberg representation. Ignoring any boundary terms, this gives
\begin{align}
\frac{d\gravham\lapsh}{dt}&=\{\gravham\lapsh,\cqadm\lapsh\}
+\frac{1}{2}\int dx dx' D_2^{ijkl}(x,x')\frac{\delta^2}{\delta\pi^{ij}(x)\delta\pi^{kl}(x')}\gravham\lapsh
\nonumber\\
&=\{\gravham\lapsh,\qmatterham\lapsh\}+\int dx N^2 D_2 \sqrt{g}
\frac{1}{2}\left(g^{ik}g^{jl}+g^{il}g^{jk}-2\beta g^{ij}g^{kl}\right)G_{ijkl}
\nonumber\\
&=\{\gravham\lapsh,\qmatterham\lapsh\}
+(d^2-\frac{d}{2})(1+\beta) D_2\int dx N^2\label{eq:grav_flux}
\end{align}
where $d$ is the number of spatial dimensions. We have assumed that the lapse and shift $N,\vec{N}$ only depend on the metric and not it's conjugate momenta. The first term
 is just the change in energy due to the back-reaction of the matter Hamiltonian, and will be compensated for by the action of $\gravham\lapsh$ on $\qmatterham\lapsh$
which results in an equal and opposite energy flux
 $-\{\gravham\lapsh,\qmatterham\lapsh\}$ into the matter degrees of freedom.  It is the second term which is due to the diffusion. Taking $\beta<1$ results in an apparent negative flux of gravitational kinetic energy. It can at least partially compensate for the heating of the matter field given in Equation \eqref{eq:heating}.
The question is a subtle one -- for example, in the weak field limit, what initially appears to be constant anomalous energy production in the gravitational degrees of freedom turns out to be a gauge artefact\cite{layton2023weak}. The dependence of Equation \eqref{eq:grav_flux} on the lapse suggests this might be the case more generally.
	
Whether this back-reaction is enough to restore equilibrium or slow the production of energy in the matter sector of the theory is an open question. It may merely be that the total Hamiltonian is conserved on average, with heating occurring in the matter degrees of freedom, and a negative flux of kinetic energy occurring in the gravitational degrees of freedom.
To what extent this would be observable has since been partially taken up in \cite{oppenheim2021gravitationally}, where it is shown that the decoherence (and heating) in the matter degrees of freedom can be made small, but only at the expense of large diffusion in the gravitational degrees of freedom. However, because gravity is so weak, the effect appears to be within current experimental bounds, although they do constrain $D_2^{ijkl}(x,x')$. %
The question then, becomes whether the anomalous heating and diffusion are within experimental bounds, while at the same time, respecting various local symmetry properties such as diffeomorphism invariance, or some weaker version. 

Related to this, is the question of whether 
  the Hamiltonian and momentum constraint is preserved in time. In the deterministic theory, this says that the total energy and momentum density is everywhere zero. These constraints are however modified in the theory considered here. Since conservation of the constraints in general relativity enforces diffeomorphism invariance, the issue of diffusion of gravitational kinetic energy, and anomalous heating, is linked with whether diffeomorphism invariance or a related local symmetry will be respected.  Whether the theory satisfies some local symmetry, then becomes a question of satisfying some potentially modified constraint equations, and whether one can satisfy them all at the same time. %
  This turns on the question of compatibility of the constraint equations i.e. whether the {\it algebroid of constraints}\cite{algebroid_foot} closes. %
  Of course in the CQ theory, the constraint is not of the form Equation \eqref{eq:badconstraint}, and the full dynamics are subtle. A fuller understanding of the algebroid of constraint equations thus remains an open issue, with some initial progress reported in \cite{UCL2022constraints} where these issues are further explored (cf. \cite{oppenheim2023covariant}).
  
There is some reason to believe 3+1 diffeomorphism invariance or something related to it can be respected, given the diffeomorphism invariant realisation of \cite{oppenheim2023covariant}, based on the trace of Einstein's equations, and using the path integral formulation of CQ dynamics\cite{oppenheim2023path}. In  \cite{oppenheim2023covariant} we also present a diffeomorphism invariant theory which gives all of Einstein's equations, although we have not proven that the dynamics is CPTP on valid initial states. The  challenge then, is to see if the realisation is consistent with experimental constraints such as those on anomalous heating and gravitational diffusion, and is ideally also renormalisable in the matter degrees of freedom. While realisations exist which are renormalisable, and realisations exist which have small amounts of anomalous heating, the question of whether a realisation exists which satisfies all these requirements remains open.\phantomsection\label{par:challenge}

The classical-quantum theory has a number of experimental signatures, some of which have since been quantified in  \cite{oppenheim2021gravitationally}. It predicts a form of gravitational decoherence, which already a number of experiments are looking for\cite{bassi2013models,schmole2016micromechanical,carney2018massive}. %
It predicts stochastic, finite sized jumps\foothide{Here, space-time, as well as $g_{ab},\pi^{ab}$ are taken to be continuous, while the jumps in phase space are not. It's possible to imagine a theory in which the metric is discrete or even space-time itself, however, 
	there is no need to go this far in the present context.} or continuous diffusion of the gravitational field which might be detectable in the lab with current technology. 
The theory also predicts diffusion which might be observable in astrophysical systems and in cosmology. The theory is highly constrained, especially if $\tau$ the parameter in Equation \eqref{eq:weiner2} is taken to be of order $\hbar$, since increasing the coherence time results in larger jumps and more diffusion. 
Finally, the locally correlated realisations considered here predict a null result to recently proposed experiments to test the quantum nature of gravity\cite{kafri2013noise,bose2017spin,marletto2017gravitationally,entanglement_foot}. 

Here, space-time is treated as fundamentally classical, but one could instead  view this theory as  a candidate for an effective theory which results from taking the classical limit of the gravitational degrees of freedom of a fully quantum theory of gravity. 	There are a number of ways of taking this partial classical limit. The extent and the conditions under which this gives the equations of motion presented here is taken up in \cite{UCLQQtoCQ}. We find that care should be taken, since an effective theory might violate our assumptions of complete positivity at short time scales, or of being time-local. Nonetheless we find that there is a regime, and a choice of parameters, such that Equation \eqref{eq:cq-dynamics} describes the consistent evolution of interacting quantum systems in the limit where a subsystem behaves classically.

 Whether the theory presented here is taken to be fundamental or not, it provides a consistent way to explore the regime where space-time is classical in a manner which does not suffer from the pathologies of the semiclassical equations.  This has been lacking in our investigation into how space-time reacts to quantum systems.  This gives hope that the ability to consistently study back-reaction effects finds application on large distance scales, in cosmology, and in black-hole evaporation.

{\bf Acknowledgements}
I would like to thank Dorit Aharonov, Robert Alicki, Sougato Bose, Lajos Diosi, Patrick Hayden, Michal Horodecki, Veronika Hubeny, David Jennings, Don Marolf, Lluis Masanes, Sandu Popescu, John Preskill, Benni Reznik, Lenny Susskind, Bill Unruh, and groups 17 and 18 of PHAS3441 (2018) for interesting discussions. I would especially like to thank 
Joan Camps, Barbara Šoda, Carlo Sparaciara, 
and Zach Weller-Davies for helpful discussion clarifying some of the points raised here. I am grateful to David Poulin for introducing me to Equation \eqref{eq:BJAKDP-dynamics} and his comments on an early version of the ADM realisation of the theory during the KITP QINFO17 program, as well as Lajos Diosi and Antoine Tilloy for drawing my attention to their work. I'd also like to thank Lydia Marie Williamson for correcting a number of typos from an earlier version of this draft. I am very grateful to the anonymous referees of PRX for their comments and attention to detail, which have helped improve this work.
This research was supported by an EPSRC Established Career Fellowship, a Royal Society Wolfson Merit Award, the National Science Foundation under Grant No. NSF PHY11-25915 and by the Simons Foundation {\it It from Qubit} Network. 

\raggedright 
\bibliography{common/refgrav2,common/refcq,common/refjono2,common/rmp12,common/pqg_v3_extras,common/footnotes_to_biblio}

\begin{thebibliography}{194}
\bibitem [{clu()}]{cluster_foot}%
  \BibitemOpen
  \href@noop {} {}\bibinfo {note} {Evolution can be local, in the sense that
  two local operators $A(x,t)$ and $B(x',t')$, commute at space-like separated
  points $\left[A(x,t),B(x',t')\right]=0$ whenever $(x-x')^2-(t-t')^2\geq 0$,
  but still violate cluster-separability, meaning that
  $\frac{d}{dt}\left(A(x,t)B(x',t)\right)\neq
  \frac{dA(x,t)}{dt}B(x',t)+A(x,t)\frac{dB(x',t)}{dt} $ and so correlations get
  destroyed or are created over space-like distances.}\BibitemShut {Stop}%
\bibitem [{mat()}]{matrix_foot}%
  \BibitemOpen
  \href@noop {} {}\bibinfo {note} {\label{ft:linear}If someone prepares a
  system in one of two states $\hat{\sigma}_0$ or $\hat{\sigma}_1$ depending on
  the value of a coin toss ($\proj{0}$ with probability $p$, $\proj{1}$ with
  probability $1-p$), then the evolution ${\cal L}$ of the system must satisfy
  $p\proj{0}\otimes {\cal L}\hat{\sigma}_0 + (1-p)\proj{1}\otimes{\cal
  L}\hat{\sigma}_1 = {\cal L}(p\hat{\sigma}_0 +(1-p)\hat{\sigma}_1)$ otherwise
  the system evolves differently depending on whether we are aware of the value
  of the coin toss. A violation of linearity further implies that when the
  system is in state $\hat{\sigma}_0$ it evolves differently depending on what
  state the system would have been prepared in, had the coin been $\proj{1}$
  instead of $\proj{0}$. This motivates our restriction to linear
  theories.}\BibitemShut {Stop}%
\bibitem [{bar()}]{barb_foot}%
  \BibitemOpen
  \href@noop {} {}\bibinfo {note} {As Barbara Šoda noted, one can generalise
  this formalism to the case where the connection between the Hilbert spaces at
  points $\z$ is non-trivial.}\BibitemShut {Stop}%
\bibitem [{car()}]{carlo_foot}%
  \BibitemOpen
  \href@noop {} {}\bibinfo {note} {A helpful suggestion by Carlo Sparaciari has
  streamlined the presentation of this section.}\BibitemShut {Stop}%
\bibitem [{alg()}]{algebroid_foot}%
  \BibitemOpen
  \href@noop {} {}\bibinfo {note} {We use the term to mean a generalisation of
  the usual constraint algebra, since we may require more than one bilinear map
  (the Poisson bracket, the commutator, a {\it CQ bracket} such as that of
  footnote \ref{ft:cqbracket}, as well as the stochastic terms of the master
  equation which are generally not bilinear
  maps)\label{ft:algebroid}.}\BibitemShut {Stop}%
\end{thebibliography}

\justifying 

\appendix

\section{Conditions for complete positivity of a CQ-map}
\label{sec:CP-conditions}

Here we extend a theorem due to 
Kossakowski\cite{kossakowski1972} for a cq-map ${\cal E}(t)$ to be completely positive on a separable Hilbert space\footnote{see also \cite{oppenheim2021gravitationally} for a more complete proof using indicator functions.}.
Let us consider all complete sets of projectors $P_r(\z)$ on the doubled Hilbert space $\mathcal{H_S}\otimes\mathcal{H_S}$ and classical degrees of freedom $\z$. 
\begin{theorem}
$\mathcal{L}$ is a generator of continuous time, completely positive,  Markovian evolution of $\psiz$ which preserves $\int d\z\tr\psiz$ if and only if for all sets of orthogonal projectors $\{P_r(\z)\}$:
\begin{align}
&\tr P_s(\z)\mathcal{L}_{\z|\z'}\otimes\id(P_r(\z')) \geq 0 & \forall r\neq s, \z,\z'
\label{eq:oldcondition}
\\
&\tr P_s(\z)\mathcal{L}_{\z|\z'}\otimes\id(P_s(\z')) \geq 0 & \forall  s,\z \neq \z'\label{eq:newcondition}
\\
&\int_\z \tr \mathcal{L}_{\z|\z'}\otimes\id(P_r(\z')) = 0 &\forall  r,\z'
\label{eq:tp-cq}
\end{align}
where ${\cal L}\psiz=\int d\z'\mathcal{L}_{\z|\z'}\psizp$
\label{thm:cqkos}
\end{theorem}
As a consequence of the above, we automatically  get 
\begin{align}
\tr\int d\z'\delta(\z-\z') P_s(\z)\mathcal{L}_{\z|\z'}\otimes\id(P_s(\z')) \leq 0 \s \forall  s,\z
\end{align}
since the probability conservation condition, Equation \eqref{eq:tp-cq} ensures that outward flux from any $P_s(\z)$ contributes negatively to the probability rate. The proof of Theorem \ref{thm:cqkos} follows from writing
\begin{align}
\mathcal{L}_{\z|\z'}:=\lim_{\dt\rightarrow 0}\frac{\mathcal{E}_{\z|\z'}(\dt)-\id\delta(\z-\z')}{\dt}
\end{align} 
where $\mathcal{E}(t)\psiz=\int d\z'\mathcal{E}_{\z|\z'}(t)\psizp$,
and using the fact that 
the map $\mathcal{E}_{\z|\z'}$ is completely positive, probability conserving map if and only if
\begin{align}
\tr P_s(\z)\mathcal{E}_{\z|\z'}\otimes\id (P_r(\z'))\geq 0 \s \forall  \{P_s(\z)\}, \{P_r(\z')\}, \z,\z'
\end{align} and 
\begin{align}
\sum_\z\tr\mathcal{E}_{\z|\z'}\otimes\id(P_r(\z'))=1\s\forall \{ P_r(\z)\},r,\z'\s.
\end{align} 
It follows that $\mathcal{E}_{\z|\z'}$ is trace non-increasing
\begin{align}
\mathcal{E}_{\z|\z'}\otimes\id (P_r(\z'))\leq\id
\end{align}
Equation \eqref{eq:newcondition} is an additional constraint on the quantum system not found in  \cite{kossakowski1972}, although if one considers the projector to also be on the classical degree of freedom, then it becomes subsumed by Equation \eqref{eq:oldcondition}. From Kraus's theorem, we know that this implies
that $\X^{\alpha\beta}(\z|\z';t)$ in Equation \eqref{eq:cq-kraus2} needs to be
positive semi-definite in $\alpha,\beta$ for all $\z,\z'$.

\section{Heisenberg representation}
\label{sec:heisenberg}

Equations \eqref{eq:cq-dynamics} and \eqref{eq:weiner2} tells us how the classical-quantum state evolves. As with both quantum and classical mechanics, there is an alternative ''Heisenberg'' picture in which the state doesn't evolve but the operators do. One can easily derive the time evolution of operators in this representation, since it follows from demanding that for all $\psiz$
\begin{align}
\frac{d}{dt}\int 
\dz
\tr \hatA_S(\z;t)\varrho_S(\z)
=
\frac{d}{dt}
\int 
\dz
\tr \hatA_H(\z;t)\varrho_H(\z)
\label{eq:StoH}
\end{align}
where the subscript $S$ indicates the Schr\"odinger representation, and $H$, the Heisenberg representation. Using this and Equation \eqref{eq:cq-dynamics} for the evolution of $\varrho_S(\z)$  leads to the Heisenberg evolution equation for the operator $\hatA(\z;t)$ over phase space
\begin{align}
\frac{d\hatA_H(\z;t)}{dt}
=&
\frac{\partial\hatA_H(\z;t)}{\partial t}
+i[\Hq(\z),\hatA_H(\z;t)]
\nonumber\\
 &+ \int \ddist 
  \rate^{\alpha\beta}(\z+\dist|\z)\L_{\beta}^\dagger
  \hatA_H(\z+\dist,t)
  \L_{\alpha}
  -
  \frac{1}{2}
\0mom
\{
\L_\beta^\dagger\L_\alpha,
\hatA_H(\z;t)
\}_+
\nonumber\\
=&
\frac{\partial\hatA_H(\z;t)}{\partial t}
+i[\Hq(\z),\hatA_H(\z;t)]
+
  \0mom\Big[\L_{\beta}^\dagger\hatA_H(\z;t)\L_{\alpha}
  -\frac{1}{2}
\{\hatA_H(\z;t),\L_{\beta}^\dagger(x)\L_{\alpha}\}_+   
\Big]
\nonumber\\
&-
\L_{\beta}^\dagger\{h^{\alpha\beta},\hatA_H(\z;t)\PB\L_{\alpha}
  \Big] + \cdots
\label{eq:Heisenberg}
\end{align}
where we have used cyclicity of operators under the trace, and integration by parts. We have included the purely classical term in $\rate^{00}$ but for deterministic evolution, it would be $-\{H(\z),\hatA_H(\z;t)\PB$. For a field theory, and in general relativity, we are usually interested in the evolution of diffeomorphism invariant operators such as
\begin{align}
\hatA_H(\z)=\int dx\sqrt{g}\hatA(\z;x)
\end{align}
where $A(\z;x)$ is a local scalar operator.
In such a case, we have
\begin{align}
\frac{d\hatA_H(\z;t)}{dt}
=&
\frac{\partial\hatA_H(\z;t)}{\partial t}
+i[\Hq(\z),\hatA_H(\z;t)]
\nonumber\\
 &+ \int dx\Big[\ddist 
  \rate^{\alpha\beta}(\z+\dist|\z;x,y)\L_{\beta}^\dagger(y)
  \hatA_H(\z+\dist,t)
  \L_{\alpha}(x)
  -
  \frac{1}{2}
\rate(\z;x,y)
\{\L_\beta^\dagger(y)\L_\alpha(x),\hatA_H(\z;t)
\}_+\Big]
\nonumber\\
=&
\frac{\partial\hatA_H(\z;t)}{\partial t}
+i[\Hq(\z),\hatA_H(\z;t)]
+\int 
  \rate\ab(\z;x,y)\Big[\L_{\beta}^\dagger(y)\hatA_H(\z;t)\L_{\alpha}(x)
  -\frac{1}{2}
\{\L_{\beta}^\dagger(y)\L_{\alpha}(x),\hatA_H(\z;t)\}_+   
\Big] dx
\nonumber\\
&-
\int dx \L_{\beta}^\dagger(x)\{h^{\alpha\beta}(\z;x),\hatA_H(\z;t)\PB\L_{\alpha}(x)
   + \cdots
\end{align}
Note that care must be taken here. The Heisenberg representation allows one to compute the change of the expectation value of an observable $\hatA$ with time. However, if the evolution is non-unitary, then it is not necessarily the case that $\langle \hatA^n(t)\rangle=\langle \hatA(t)\rangle^n$. For example, if there is dispersion, than typically $\langle \hat{p}^2\rangle-\langle \hat{p} \rangle^2$ will increase with time. Thus, knowing $\hatA(t)$ does not necessarily allow one to deduce the expectation value other moments of the operator.

\end{document}